%% file: master.tex
\documentclass[sigconf]{acmart}

\usepackage{url}

\usepackage{pifont}
\usepackage{color}
\usepackage{colortbl}
\usepackage{multirow}
\usepackage{graphicx}
\usepackage{array}
\usepackage{enumerate}
\usepackage{algpseudocode}

\usepackage{hhline}
\usepackage{makecell}

\usepackage[utf8]{inputenc}
\usepackage{listings}
\usepackage{subcaption}
\makeatletter
\newcommand{\removelatexerror}{\let\@latex@error\@gobble}
\makeatother

\definecolor{CodeGray}{RGB}{220,220,220}
\definecolor{ForestGreen}{RGB}{34,139,34}
\definecolor{codegreen}{rgb}{0,0.6,0}
\definecolor{codegray}{rgb}{0.5,0.5,0.5}
\definecolor{codepurple}{rgb}{0.58,0,0.82}
\definecolor{codegreen}{rgb}{0,0.6,0}
\definecolor{backcolour}{rgb}{0.95,0.95,0.92}

\lstdefinestyle{mystyle-1}{
    commentstyle=\color{codegreen},
    keywordstyle=\color{codepurple},
    numberstyle=\tiny\color{codegray},
    stringstyle=\color{codepurple},
    basicstyle=\footnotesize,
    breakatwhitespace=false,         
    breaklines=true,                 
    captionpos=b,                    
    keepspaces=true,                 
    numbers=none,                    
    numbersep=5pt,                  
    showspaces=false,                
    showstringspaces=false,
    showtabs=false,  
    frame=lines,
    escapeinside=``,            
    tabsize=2
}

\lstdefinestyle{mystyle-2}{
    commentstyle=\color{codegreen},
    keywordstyle=\color{codegreen},
    numberstyle=\tiny\color{codegray},
    stringstyle=\color{codepurple},
    basicstyle=\footnotesize,
    breakatwhitespace=false,         
    breaklines=true,                 
    captionpos=b,                    
    keepspaces=true,                 
    numbers=left,                    
    numbersep=5pt,                  
    showspaces=false,                
    showstringspaces=false,
    showtabs=false,  
    frame=single,            
    tabsize=2
}

\usepackage{cleveref}
\crefname{section}{§}{§§}

\usepackage{setspace}
\usepackage{xspace}
\newcommand{\name}{\textsc{RTron}\xspace}

\newenvironment{packeditemize}{
\begin{list}{$\bullet$}{
\setlength{\labelwidth}{8pt}
\setlength{\itemsep}{0pt}
\setlength{\leftmargin}{\labelwidth}
\addtolength{\leftmargin}{\labelsep}
\setlength{\parindent}{0pt}
\setlength{\listparindent}{\parindent}
\setlength{\parsep}{0pt}
\setlength{\topsep}{3pt}}}{\end{list}}

\makeatletter  
\newif\if@restonecol  
\makeatother

\usepackage[linesnumbered,ruled,vlined]{algorithm2e}

\begin{document}
\title{Risk Analysis and Policy Enforcement of Function Interactions in Robot Apps} 

\author{\rm Yuan Xu\textsuperscript{1}, Tianwei Zhang\textsuperscript{2}, Yungang Bao\textsuperscript{1}\\
\textsuperscript{1}ICT, CAS  \quad \quad
\textsuperscript{2}Nanyang Technological University}

\begin{abstract}
Robot apps are becoming more automated, complex and diverse. An app usually consists of many functions, interacting with each other and the environment. This allows robots to conduct various tasks. However, it also opens a new door for cyber attacks: adversaries can leverage these interactions to threaten the safety of robot operations. Unfortunately, this issue is rarely explored in past works. 

We present the \emph{first} systematic investigation about the function interactions in common robot apps. First, we disclose the potential risks and damages caused by malicious interactions. We introduce a comprehensive graph to model the function interactions in robot apps by analyzing 3,100 packages from the Robot Operating System (ROS) platform. From this graph, we identify and categorize three types of interaction risks. Second, we propose \name, a novel system to detect and mitigate these risks and protect the operations of robot apps. We introduce security policies for each type of risks, and design coordination nodes to enforce the policies and regulate the interactions. We conduct extensive experiments on 110 robot apps from the ROS platform and two complex apps (Baidu Apollo and Autoware) widely adopted in industry. Evaluation results indicated \name can correctly identify and mitigate all potential risks with negligible performance cost. To validate the practicality of the risks and solutions, we implement and evaluate \name on a physical UGV (Turtlebot) with real-word apps and environments.
\end{abstract}

\maketitle

\input{Tex/1-Introduction}

\input{Tex/2-Background}
\input{Tex/3-AppAnalysis}
\input{Tex/4-RiskAnalysis}
\input{Tex/5-Methodology}

\input{Tex/6-RTron}
\input{Tex/7-Evaluation}
\input{Tex/8-RelatedWork}
\input{Tex/9-Conclusion}








\bibliographystyle{IEEEtran}
\bibliography{refs}

\appendix
\input{Tex/Appendix}

\end{document}

%% file: Tex/1-Introduction.tex
\section{Introduction}
\label{sec:1-introduction}
The robotics technology is rapidly integrated into every aspect of our life. 
Different types of robots and applications
were designed to assist humans with many dangerous or tedious jobs.
A robot app usually consists of multiple processes (a.k.a. nodes), with each one focusing on one specific function, e.g., localization, path planning. They interact with each other to complete the end-to-end task.


To ease the development of robot apps, 
many companies expose interfaces of massive functions for their products (e.g. Ford Open XC \cite{openxc}, Dji Onboard SDK \cite{onboard-sdk}, UR Application Builder \cite{ur-app-builder}). Developers can then use these functions to create new apps. Alternatively, public platforms are introduced, where functions are developed in a crowd-sourcing manner by third-party developers and distributed through the open-source function markets. The most mainstream platform is the Robot Operating System (ROS) \cite{ros}, which provides thousands of open-source robot functions. 
Functions from this platform have been widely adopted in the research community and many commercial products, such as Dji Matrice 200 drone \cite{onboard-sdk}, PR2 humanoid \cite{ros-pr2} and ABB manipulator \cite{ros-abb}.

However, these functions can be the Achilles' Heel of robot apps, threatening the safety of robot operations. There are two reasons that facilitate this hazard. (1) Public platforms like ROS allow third-party developers to share their functions. Different from other well-developed app stores (e.g., mobile devices \cite{android-store,ios-store}, PCs \cite{windows-store,ubuntu-store,mac-store}, IoT \cite{smarthings,homekit,weave}), the ROS platform does not enforce any security inspection over the submitted code. An adversary can easily upload malicious functions to the platform for users to download. This threat is practical but severe: \emph{we successfully demonstrate the feasibility of uploading malicious functions to the ROS platform and compromising any robot apps built upon them} (\cref{sec:7-4-usecase} and Appendix \ref{sec:D-attack-impl}). (2) Function nodes in a robot app have dynamic and frequent interactions with each other and the physical environment. Even one malicious node can affect the states and operations of the entire app, leading to severe privacy breach and physical damages \cite{ros-scan,sec-analysis2}.
For instance, Chrysler Corporation recalled 1.4 million vehicles in 2015 due to a software vulnerability in its Uconnect dashboard computers \cite{chrysler}. An adversary could exploit it to hack into a jeep remotely and take over the dashboard functions. 


To ensure the safety of robot apps, it is critical to protect the interactions among various functions inside the apps. We are interested in two questions: \emph{What potential risks and security incidents can a malicious interaction bring?}
\emph{How can we detect and mitigate malicious interactions?} 
Unfortunately, there are currently few studies focusing on the interactions in robot apps. 
Security analysis of interactions in IoT systems have been explored \cite{iot-sift,iot-salus,iot-iotMon,iot-iotGuard,iot-soteria,iot-iostSan,iot-homeGuard,iot-menShen,iot-autoTap,iot-iRuler}. As robot apps have more complex and distinct features, it is hard to apply the above methods to the robot ecosystem, as discussed in \cref{sec:8-related-work}. 

In this paper, we present the \emph{first} study to explore the function interactions in common robot apps from the perspective of security and safety. We make three contributions to answer the above two questions. First, we introduce a comprehensive interaction graph to model all the interactions in robot apps. Although numerous apps have been implemented for various robot devices and tasks, there are still no systematic summaries about the characteristics of function node interactions. To achieve this, we implement a rule-based tool to automatically analyze 3100 packages from the ROS platform, categorize them into 17 types and build a graph to cover all possible interactions. 
We also select 110 popular robot apps from the ROS platform to verify our interaction graph. 
This model lays a foundation for our risk analysis and mitigation in this paper, and also aims to enhance people's understanding about the characteristics of robot apps for other purposes in the future. 

Second, we analyze potential risks from those interactions based on our graph model. We classify these risks into three types. (1) \emph{General Risk}: it happens when multiple function nodes share same states, and malicious nodes attempt to compromise the states by sending wrong messages. (2) \emph{Robot-Specific Risk}: this is caused by the conflict between the robot's velocity and the frame rate of the image recognition function. (3) \emph{Mission-Specific Risk}: this refers to the violation of users' expectation regarding the safe and secure behaviors of the robot system. We provide detailed analysis and examples to show the possible consequences of each risk. 

Third, we introduce \name, a novel system to detect and mitigate risks caused by suspicious interactions in robot apps. The core of \name is a set of \emph{coordination nodes}, which are used to regulate the interactions and enforce security policies. We design a coordinate node with some security policies to mitigate each type of risks.
Specifically, \name includes two stages. At the development stage, it generates the interaction graph from the source code of the robot app, and help \emph{developers} discover all high-risk function nodes, which may trigger potential malicious interactions. Based on the generated risk information, \name deploys \emph{coordination nodes} along with these high-risk function nodes. This is achieved without changing the original function node.  
At the operation stage, \name deploys a security service to keep monitoring all the information from the coordination nodes. A visualized interface is provided to \emph{end users} to observe the high-risk interactions. If a risk occurs, the corresponding coordination node will enforce the desired policy configured by users during the app launch to mitigate it.


We conduct extensive experiments to evaluate the effectiveness, efficiency and practicality of \name. (1) We select 110 robot apps from the ROS platform, covering 24 robots of 4 types. \name can correctly identify all potential risks from three types of vulnerable interactions, with negligible overhead at both the offline and online stages. (2) We perform large-scale evaluations on more complex and practical robot apps: we select 2 apps from the ROS platform for the home and autorace scenarios, each containing 10 functions to perform 6 tasks; we also deploy 2 self-driving apps (Autoware \cite{app-autoware} and Apollo \cite{app-apollo}), which are widely adopted in the autonomous vehicle industry. \name successfully identifies 198 high-risk interactions in these 4 apps, and mitigates them promptly and effectively. (3) We demonstrate a practical end-to-end attack, from uploading malicious functions to the public ROS platform, developing vulnerable robot apps, to compromising a physical robot (Turtlebot UGV) and environment. We show this threat can be eliminated by \name. 

%% file: Tex/2-Background.tex
\section{Background \& Threat Model}
\label{sec:2-background}

\subsection{Interaction in Robot Apps}
\label{sec:2-app}

\begin{figure}[tb]
\centering
\includegraphics[width=1.0\linewidth]{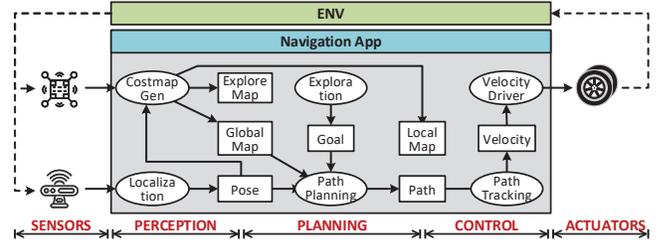}
\vspace{-20pt}
\caption{An example of the navigation app.}
\vspace{-15pt}
\label{fig:2-robot-app}
\end{figure}

Robot apps run on the embedded computer of a robot device to interpret sensory data collected from the environment, and make the corresponding action decisions. The workflow of a robot app can be represented as a directed graph, where each node represents a certain function, and edges represent the dependencies of the functions in this app. Figure \ref{fig:2-robot-app} shows a navigation app as an example. This robot app is composed of three major processing stages \cite{handbook}: (1) \emph{Perception}: the robot extracts estimated states of the environment and the device from raw sensory data. It uses the \texttt{Localization} node to determine the device position, and the \texttt{CostmapGen} node to model the surroundings. (2) \emph{Planning}: the robot determines the long-range actions. It uses the \texttt{Path Planning} node to find the shortest path, and the \texttt{Exploration} node to search for all accessible regions. The \texttt{Exploration} node also exposes an external service for users to launch a navigation mission. (3) \emph{Control}: the robot processes the execution action and forwards these motions to the actuators. It uses the \texttt{Path Tracking} node to produce velocity commands following the planned path, and the \texttt{Velocity Driver} node to convert the velocity to instructions for the motor to drive the wheels.

One big feature of robot apps is the high interactions among various function nodes in the workflow. Based on the triggered events, the interactions can be classified into two groups: 

\noindent
\textbf{Direct interaction (solid line). }
This denotes the interaction between two functions (ellipses), which are directly connected in the workflow and sharing common robot states (squares). Robot states are defined as the collection of all aspects and knowledge of the device that can impact future behaviors \cite{pr}, e.g., position, orientation, explored maps. The computation of one function can change some robot states, which will affect the computation of another function. For instance, in Figure \ref{fig:2-robot-app}, the action of \texttt{Path Planning} is triggered by the event that \texttt{Localization} generates the robot's current position and orientation. Then the two nodes have direct interaction over the robot states of position and orientation.

\noindent
\textbf{Indirect interaction (dotted line). }
This refers to the dependency of two functions, which are not connected in the workflow, but can interact with each other via the environmental context.
One node in the app can issue actions to change the environmental context (e.g., obstacles, space, etc.), which will further influence another node. In the navigation app, the functions in the \emph{Control} stage generate commands to control the robot to change the physical environment. This triggers the functions in the \emph{Perception} to conduct new computations. For instance, the map created by the \texttt{CostmapGen} function depends on the action from the \texttt{Path Tracking} function. As a result, these two function nodes are indirectly interacted, although they are not directly connected in the workflow.

\subsection{Robot App Platform}
\label{sec:2-ros}
In robotics, the most popular app platform is Robot Operating System (ROS) \cite{ros}. Both the research community and industry widely adopt ROS as the foundation or the testbed for their apps, such as Dji Matrice 200 drone \cite{onboard-sdk}, PR2 humanoid \cite{ros-pr2} and ABB manipulator \cite{ros-abb}. In this paper, we mainly focus on the ROS platform. Our methods and conclusions can be generalized to other platforms as well. 

The ROS platform offers two kinds of services. First, it provides \emph{robot core libraries}, which act as the middleware between robot apps and hardware. These core libraries support hardware abstraction, message passing mechanisms and device drivers for hundreds of sensors and motors. Second, the ROS platform maintains thousands of \emph{robot code repositories} (a.k.a. repos) for distributed version control, code management and sharing.
A repo commonly includes one or more packages maintained on the hosting site (e.g. GitHub, BitBUcket, GitLab). The developers can add their repos to the ROS platform through sending a \textit{pull} request to the ROS maintainer. If it succeeds, both the repos and included packages can get specific indexes for other developers to download and use. Detailed information regarding the ROS platform and repos can be found in Appendix \ref{sec:repo-func-pkg}. This work mainly targets this service, and shows that untrusted repos from the ROS platform can significantly threaten the robot apps built from them.

\begin{figure}[tb]
\centering
\includegraphics[width=1.0\linewidth]{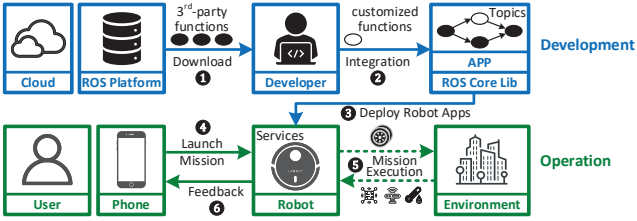}
\vspace{-20pt}
\caption{The lifecycle of robot app development (blue parts) and operation (green parts).}
\label{fig:2-ros-arch}
\vspace{-10pt}
\end{figure}

Figure \ref{fig:2-ros-arch} illustrates the key concepts and components in the lifecycle of robot app development and operation. First, the design of the app is decomposed into several necessary functions. Among them core functions (white ellipses) need to be customized by the developer, while non-core functions (black ellipses) can be downloaded from ROS code repos (\ding{182}). Then the developer uses ROS core libraries to organize these functions as an app workflow (\ding{183}) and deploys the app to the robot (\ding{184}). Each function is abstracted as a ROS node and connected with others through ROS \emph{Topics}. The ROS topics are many-to-many named buses that store the robot or environment states. Each topic is implemented by the \emph{publish-subscribe} messaging protocol: some nodes can subscribe to a topic to obtain relevant data, while some nodes can publish data to a topic. 

The robot communicates with end users through ROS \emph{Services}. 
The ROS services are a set of interfaces of the robot app exposed to end users. Each service is implemented by the \emph{Remote Procedure Call} (RPC) protocol and allows users to launch tasks or adjust function parameters. 
Once the robot receives a mission from the user's phone (\ding{185}), it executes the mission and interacts with the surrounding environment at runtime (\ding{186}). The user will receive the notification from the robot when all tasks are completed (\ding{187}). 

\subsection{Threat Model and Problem Scope}
\label{sec:2-threat}
In this paper, we consider a threat model where some nodes of a robot app are untrusted. Those adversarial nodes aim to compromise the robot's operations, forcing it to perform dangerous actions. This can result in severe security and safety issues to machines, humans and environments \cite{ros-vulnerable,robot-vulnerable,hackingrobots}. 

This threat model is drawn from three observations. First, the ROS platform is open for everyone to upload and share their code repos. Different from app stores of other ecosystems \cite{android-store,ios-store,windows-store,ubuntu-store,mac-store,smarthings,homekit,weave}, the ROS platform \emph{does not have any security check over the submitted code}. As a result, an adversarial developer can insert malicious code to a repo and publish it to the ROS platform for other users to download. This has been highlighted in the design document of ROS2 Robotic Systems Threat Model \cite{ros-3rd-party-threat}: ``\emph{third-party components releasing process create additional security threats (third-party component may be compromised during their distribution)}''.  We also confirm the feasibility and practicality of this threat by successfully publishing malicious packages to the ROS platform, in an end-to-end attack demonstrated in \cref{sec:7-4-usecase} and Appendix \ref{sec:D-attack-impl}.
Second, the quality of third-party function code is not guaranteed. A lot of functions in the ROS platform are in a lack of coding standards or specifications. They may also contain software bugs that can be exploited by an adversary to compromise the nodes at runtime \cite{ros-vulnerable,ros-assessment}. 
By inspecting the latest commit logs in the Robot Vulnerability Database \cite{rvd}, 17 robot vulnerabilities and 834 bugs (e.g., no authentication, uninitialized variables, buffer overflow) were discovered in the repos of 51 robot components, 37 robots and 34 vendors in the ROS platform. Most of them are still not addressed yet.
Third, the high interactions among nodes in a robot app can amplify the attack damage. If an adversary controls one node, it is possible that he can affect other nodes directly or indirectly, and then the entire app. The existence of untrusted nodes can also cause data races or deadlocks when the interaction synchronization is not well handled. 

Given this threat model, our goal is to design a methodology and system, which can identify and mitigate the safety risks caused by the malicious nodes inside robot apps. We focus on the protection of node interactions (both direct and indirect) instead of the operation of individual nodes. We further assume the underlying OS and ROS core libraries are trusted: the operational flow and data transmission are well protected, and the isolation scheme is correctly implemented so the malicious nodes are not able to hijack the honest ones or the privileged systems. How to enhance the security of the ROS core libraries \cite{ros-authen-1,ros-authen-2,ros-authen-3,ros-vulnerable} and mitigate vulnerabilities from networks \cite{crypto-1,crypto-2,crypto-3,dos-2}, sensors \cite{spoof-gps1,spoof-gps2,spoof-gps3,spoof-gps4, spoof-gps5, spoof-lidar-1, spoof-lidar-2, spoof-optical, spoof-gyroscopic-1, spoof-gyroscopic-2, spoof-gyroscopic-3, spoof-optical, inject-magnetic}, actuators \cite{spoof-actuator,spoof-signal} and controllers \cite{sec-analysis} are orthogonal to our work.

%% file: Tex/3-AppAnalysis.tex
\section{Function Interaction Analysis}
\label{sec:3-app-analysis}

In this section, we aim to draw an interaction graph to model all the functions in the ROS platform and their communications. Up to the date of writing, the ROS platform contains 941 repos with around 3,100 packages. A function is typically composed of multiple packages, while a repo is usually developed for one specific function. Hence, we first implement an automatic tool to categorize these repos based on the function type they can achieve\footnote{There are a few repos which can realize more than one functions. Our tool splits them into sub-repos with each one focusing on one function, and then categorizes them.} (\cref{sec:3-1-model}). Then we build an interaction graph based on this categorization (\cref{sec:3-2-graph}), which is further verified by common robot apps (\cref{sec:3-3-app}).

\subsection{Categorization of Robot Functions}
\label{sec:3-1-model}


We first build an automatic tool to perform large-scale analysis on all the repos from the ROS platform. Past works adopted Natural Language Processing and code dependency clustering techniques to analyze smartphone and home-based IoT apps \cite{nlp-1,nlp-2,nlp-3,nlp-4,nlp-5,nlp-6,nlp-7}. However, it is challenging to apply these techniques to the ROS apps. First, a large portion of repos have poor code quality and the descriptions are not well documented, which can hardly reflect the characteristics of the functions. According to our analysis, 340 out of 941 repos do not provide the function descriptions in the documents, while 14 repos even use Japanese or German for the description.
Second, the dependencies across repos do not provide accurate or useful information for clustering. On one hand, most repos achieve simple functions (e.g. localization, teleoperation) without importing any other repos. On the other hand, some common repos are imported by various repos without any connections. For instance, \texttt{cv\_bridge} calls the OpenCV library to process images. It can be imported by the recognition,  QR-based localization ror vision-based mapping.

Instead, we adopt the rule-based approach to analyze and categorize repos based on their names, manifest files and related documents (e.g. README file or Wiki page). We adopt the Stanford TokensRegex \cite{tokens-regex} to implement such an automatic tool.  Detailed implementation is described in Appendix \ref{sec:A-auto-analysis}. In our evaluation, this tool can successfully identify the function types of 88.52\% ROS repos. The rest repos (11.48\%) which have too blur language features to be identified are then analyzed manually, with very minor effort. We believe our tool can adapt to future ROS repos as well.

Table \ref{table:repository} illustrates our analysis results about all function types of repos under the ROS1 kinetic version in the platform: 941 repos are spit into 1135 ones as some repos implement two or more functions. Then they are categorized into 17 function types in five domains.

\begin{table}[tb]
\caption{Categorization of repos in the ROS platform.}
\vspace{-10pt}
\centering
\resizebox{\linewidth}{!}{
\begin{tabular}{|l|l|l|l|}
\hline
\rowcolor{gray!40} 
\textbf{Domain} & \textbf{Function Type} & \textbf{\# of repos} & \textbf{Example repos} \\
\hline
\multirow{4}{*}{\makecell[l]{Perception \\(17.6\%)}} 
		& Preprocessing 	& 84 (7.4\%) 	& lidar\_camera\_calibration \\
	\cline{2-4}
    	& Localization  	& 35 (3.1\%) 	& slam\_karto \\
    \cline{2-4}
    	& Mapping 	    	& 31 (2.7\%)   	& homer\_mapping \\
    \cline{2-4}
    	& Recognition   	& 50 (4.4\%) 	& jsk\_recognition \\
\hline
\multirow{2}{*}{\makecell[l]{Planning \\(10\%)}}
    	& Path Planning  	& 103 (9.1\%) 	& asr\_ftc\_local\_planner \\    
    \cline{2-4}
    	& Goal Planner   	& 11 (1\%)  	& behaviortree\_planner \\
\hline
\multirow{4}{*}{\makecell[l]{Control \\(10\%)}}
		& Path Tracking  	& 68 (6\%) 	& teb\_local\_planner\_tutorials \\
	\cline{2-4}
		& Teleoperation 	& 30 (2.6\%) 	& joy\_teleop \\
	\cline{2-4}
		& Speech Generation & 9 (0.8\%)  	& homer\_tts \\
	\cline{2-4}
		& Switcher 			& 6 (0.5\%)  	& iot\_bridge \\
\hline
\multirow{4}{*}{\makecell[l]{Drivers \\(18.4\%)}}  
		& Mobile 			& 38 (3.3\%) 	& ackermann\_controller \\
	\cline{2-4}
		& Manipulator 		& 37 (3.3\%) 	& agile\_grasp \\
	\cline{2-4}
		& Speaker 			& 6 (0.5\%) 	& xbot\_talker \\
	\cline{2-4}
		& Sensors 			& 128 (11.3\%) 	& xsens\_driver \\
\hline
\multirow{3}{*}{\makecell[l]{Others \\(44\%)}} 
		& Visualization 	& 169 (14.9\%)  	& rqt\_reconfigure \\
	\cline{2-4}
		& Support       	& 111 (9.8\%)  		& aws\_ros1\_common \\
	\cline{2-4}
		& Extension     	& 219 (19.3\%) 		& ros\_pytest \\
\hline
\end{tabular}}
\label{table:repository}
\vspace{-15pt}
\end{table}




The first three domains include main functions for computing the robot tasks. 
In the \emph{Perception} domain, the \texttt{Preprocessing} function has the largest proportion, which converts the raw sensory data or calibrates multiple data to the desired representation. Then the processed data are transmitted to the \texttt{Localization} and \texttt{Mapping} functions to form the knowledge of navigation, or transmitted to the \texttt{Recognition} function to generate the corresponding information. 
In the \emph{Planning} domain, functions are used to parse a high-level task to a set of low-level actions based on the knowledge from the environment. For example, the \texttt{Path Planning} function receives the coordinate of the destination and computes a collision-free path to reach that position. The \texttt{Goal Planner} function 
encapsulates each action into a basic unit and defines the corresponding logic execution sequence. The planned actions are then forwarded to the functions in the \emph{Control} domain to drive the actuators. 
The actuators can be commonly classified into three categories: wheels or rotors in mobile robots, manipulator and speaker. The first two actuators are driven by the \texttt{Path Tracking} function, which generates adaptive velocities to control these robots navigating in the real world. The \texttt{Speech Generation} function creates audios for the speaker to respond to users' requests. This function also includes packages from many cloud providers.
The \texttt{switcher} function is used to switch on/off certain actuators for some specific tasks. 

In addition to the three main domains, we also identify two more categories. In the \emph{Drivers} domain, the functions allow the robot to interface with various actuators.
The \emph{Others} domain has the largest number of repos. Specifically, the \texttt{Visualization} function provides a GUI plugin for users to control the robot. 
The \texttt{Support} function builds a bridge between ROS and many third-part platforms, making the robot apps compatible with AI frameworks, 
mobile apps, 
and cloud services. 
The \texttt{Extension} function provides the wrappers of core libraries in ROS to ease the robot app development.

\begin{figure*}[tb]
\centering
\includegraphics[width=0.95\linewidth]{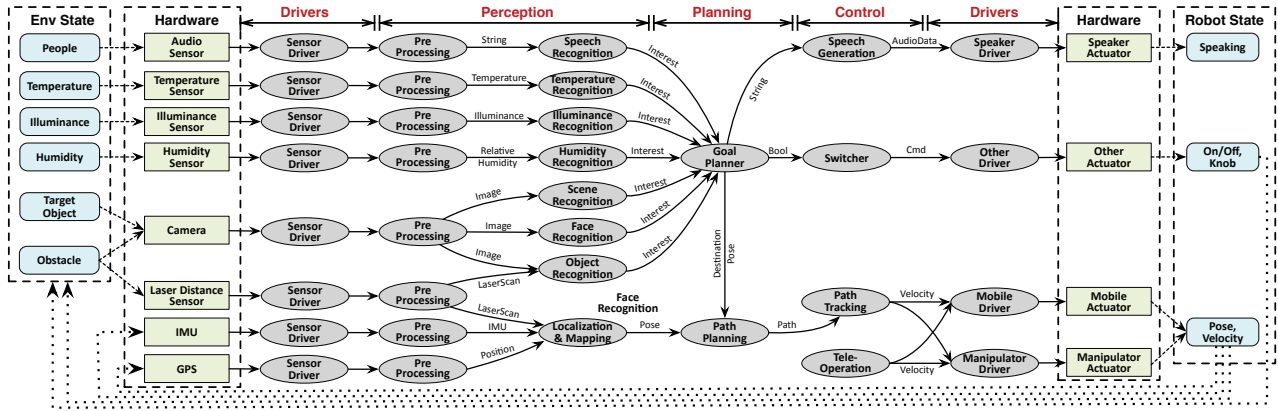}
\vspace{-10pt}
\caption{The interaction graph of robot functions.}
\label{fig:3-graph}
\vspace{-10pt}
\end{figure*}

\begin{table*}[tb]
\caption{Analysis of open-source robot apps from the ROS showcase website \cite{ros-robot}.}
\vspace{-5pt}
\centering
\resizebox{0.99\linewidth}{!}{
\begin{tabular}{|l|l|l|l|l|}
\hline
\rowcolor{gray!40} 
\textbf{App Categories} & \textbf{\# of apps} & \textbf{Robot Type} & \textbf{Set of function nodes} & \textbf{Example Robot} \\
\hline
Remote Control & 23 (20.8\%) & MB, MM, HR, MAV & Teleoperation+Mobile/Manipulator Driver & Caster \\
\hline
Panorama & 2 (1.8\%) & MB & Preprocessing & Turtlebot3 \\
\hline
2D/3D Mapping & 8 (7.3\%) & MB & Preprocessing+Mapping+Teleoperation+Mobile Driver & Xbot \\
\hline
Navigation & 22 (20\%) & MB, MM, MAV & \makecell[l]{Preprocessing+Localization+Path Planning+Path Tracking+Mobile Driver} & Tiago++ \\
\hline
SLAM & 11 (10\%) & MB & \makecell[l]{Preprocessing+Localization+Mapping+Path Planning+Path Tracking+Mobile Driver} & Roch \\
\hline
Exploration & 5 (4.5\%) & MB & \makecell[l]{Preprocessing+Localization+Mapping+Goal Planner+Path Planning+Path Tracking+Mobile Driver} & Turtlebot2 \\
\hline
Follower & 8 (7.3\%) & MB & Preprocessing+Recognition+Mobile Driver & Magni Silver \\
\hline
Manipulation & 8 (7.3\%) & MM & \makecell[l]{Preprocessing+Localization+Path Planning+Path Tracking+Manipulator Driver} & LoCoBot \\
\hline
Face/Person Detection & 8 (7.3\%) & MB, MM, MAV & Preprocessing+Recognition & ARI \\
\hline
Object/Scene Detection & 5 (4.5\%) & MM & Preprocessing+Recognition & Tiago \\
\hline
Object Search & 1 (1\%) & MM & \makecell[l]{Preprocessing+Localization+Recognition+Goal Planner+Path Planning+Path Tracking+Mobile Driver} & ROSbot 2.0 PRO \\
\hline
Gesture Recognition & 3 (2.7\%) & HR, MAV & Preprocessing+Recognition+Manipulator Driver & COEX Clover \\
\hline
Voice Interaction & 5 (4.5\%) & MB, HR & \makecell[l]{Preprocessing+Recognition+Speech Generation+Speaker/Mobile/Manipulator Driver} & Qtrobot \\
\hline
Autonomous Driving & 1 (1\%) & MB & \makecell[l]{Preprocessing+Localization+Recognition+Goal Planner+Path Planning+Path Tracking+Mobile Driver} & Turtlebot3 \\
\hline
\end{tabular}}
\label{table:apps}
\vspace{-5pt}
\end{table*}

\subsection{Building an Interaction Graph}
\label{sec:3-2-graph}
With the above categorization, we analyze the message types between different repos, and generate an abstract interaction graph, as shown in Figure \ref{fig:3-graph}.  
From the figure, we can observe that functions in \emph{Perception}, \emph{Planning} and \emph{Control} domains constitute all the computational nodes (the gray ellipse) in the interaction graph\footnote{According to the sensor types supported by ROS \cite{ros-sensors}, we further split the \texttt{Recognition} function into several recognition subfunctions, e.g. temperature/illuminance/humidity recognition.}. 

As discussed in \cref{sec:2-app}, function nodes share robot states through direct interactions. The robot states are generally estimated values the function node computes based on the sensory data. We use a solid arrow to represent the direct interaction between two nodes, which is implemented in a topic pattern. Specifically, two functions are connected if the source node publishes message types which are subscribed by the destination node. The message type published/subscribed by adjacent nodes is labeled above the arrow.
For example, the published message type of the \texttt{Path Planning} function is `nav\_msgs/Path', which is the same as the subscribed message type of \texttt{Path Tracking} function. Thus, a connection appears from the \texttt{Path Planning} to \texttt{path Tracking}.

In addition, function nodes also have indirect interactions via the physical environment. The action commands generated from the function nodes in the \emph{Control} domain can alter the robot's surrounding environment, which will also affect some function nodes as they need to re-estimate the latest robot states and determine the actions at the next moment. We use dotted arrows to represent such indirect interactions caused by the changes of environmental contexts. For example, after the \texttt{Path Tracking} node generates velocity and drives the robot to a new context, the obstacle's position relative to the robot has also changed. Then the function nodes in \emph{Perception} need to update the robot states (e.g. map and pose). If an obstacle is added to an unknown map, the \texttt{Path Planning} node may need to re-plan a new path.

With those direct and indirect interactions, Figure \ref{fig:3-graph} gives a complete closed-loop robot operation. Specifically, function nodes in the \emph{Perception} domain receive physical environmental information from sensors and convert it to robot internal estimated states (e.g. pose). Function nodes in the \emph{Planning} domain generate long-term plans (e.g. path) based on robot's knowledge. All data flows converge to the function nodes in the \emph{Control} domain, which output action commands (e.g. velocity) to control related actuators. Then the actuators can alter the surrounding environment, and force the function nodes in the \emph{Perception} to repeat the above procedure.

\subsection{App Analysis}
\label{sec:3-3-app}
Our interaction graph is built from the analysis of individual function nodes. We verify its comprehensiveness and correctness using end-to-end robot apps. 
We collect 110 robot apps from the ROS showcase website \cite{ros-robot}, covering 24 different robots including mobile base (MB), mobile manipulator (MM), micro aerial vehicle (MAV) and humanoid robot (HR). Table \ref{table:apps} summarizes the categories of these apps, numbers and the applicable robot types. All these apps can be illustrated using our interaction model, as shown in the fourth column. Detailed descriptions of the commonly used apps can be found in Appendix \ref{sec:robot-app-describe}.




Note that most of these apps analyzed above are designed for the research purpose, e.g., algorithm analysis and function testing. An industrial/commercial app needs to perform more complex tasks with a variety of unexpected emergencies in the real world. As a result, it is common to integrate multiple research apps for better robustness and functionality. For example, the Navigation app cannot recognize traffic lights and handle emergencies (e.g., intruders, accidents). We need to integrate Object/Scene Recognition to recognize these conditions and take corresponding safe operations. The combinations of these apps can still be modeled using our interaction graph, as we will demonstrate in \cref{sec:7-evaluation}. 

%% file: Tex/4-RiskAnalysis.tex
\section{Risk Analysis}
\label{sec:4-risk-analysis}
\begin{figure}[tb]
\centering
\includegraphics[width=0.9\linewidth]{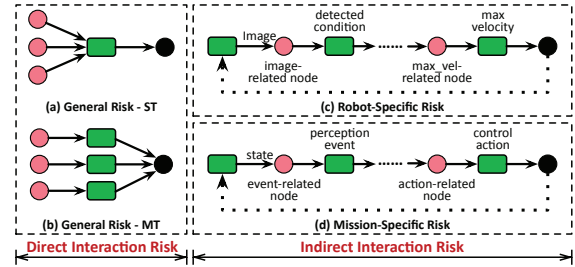}
\vspace{-10pt}
\caption{Three types of interaction risk.}
\label{fig:4-risk-type}
\vspace{-10pt}
\end{figure}

We analyze safety risks caused by malicious function nodes and interactions. We classify these risks into three categories (Figure \ref{fig:4-risk-type} and Table \ref{table:policy}). 
We describe how each risk can incur unexpected behaviors to threaten the robot safety. 

\subsection{General Risk (GR)}
GR is caused by a direct interaction. It occurs when multiple function nodes share the same robot states. If one node is malicious, it can intentionally change the robot states to wrong values to affect the robot operation. Based on the interaction graph, there are two conditions to trigger the GR. First, two or more function nodes are connected to the same successor node, and at least one of them is untrusted. Second, the transmitted message types among the above function nodes need to be the same. This guarantees that all these nodes share the same robot state through the direct interaction. 

According to the number of topics, GR can be further divided into two types. (1) General Risk with Single Topic (GR-ST): multiple high-risk nodes publish to one same topic, subscribed by the successor node (Figure \ref{fig:4-risk-type}a). (2) General Risk with Multiple Topics (GR-MT): both the indegree and outdegree of the topic are equal to 1. There can be multiple parallel topics with the same message type subscribed by the successor function (Figure \ref{fig:4-risk-type}b).

\subsection{Robot-Specific Risk (RSR)}
RSR happens in an indirect interaction, due to the conflict behaviors related to the robotic mobility characteristic. This mobility feature requires the robot to recognize real-time environment conditions (e.g. obstacle avoidance, traffic light) and react to them promptly. 
The robot's maximal velocity is determined by its reaction time, which further depends on two factors \cite{mavbench,autovehicle}. The first factor is the processing time for collision avoidance, which is the end-to-end latency from obstacle detection to velocity control. The second factor is the frame rate of the \texttt{Image Recognition} function. The faster the robot is, the larger frame rate this function requires to respond to the rapid changes of the environment. In this paper, we only focus on the second factor as the processing latency is the safety issue of the internal function node (i.e. \texttt{Path Tracking}) rather than the interaction between two nodes. 

Figure \ref{fig:4-risk-type}c shows the mechanism of RSR. There are two types of high-risk function nodes: (1) the image-related node is used to understand the current detected conditions through image recognition. (2) The max\_vel-related node outputs the maximal velocity value to the corresponding topic based on the current condition. These two nodes affect each other via an indirect interaction (dotted line). The maximal velocity and image frame rate should satisfy certain conditions to guarantee the robot can function correctly. If either node is malicious and produces anomalous output (too large maximal velocity or too small frame rate), the requirement can be compromised, bringing catastrophic effects in some tasks.

\subsection{Mission-Specific Risk (MSR)}
MSR refers to the violation of users’ expectations regarding the safe and secure behaviors of a robot system. It exists in the indirect interaction between an event-related node and action-related node (Figure \ref{fig:4-risk-type}d), when there are conflicts between them, regulated by some scenario-specific rules. Although some GRs and RSRs may also lead to the violation of these rules, the causes and mitigation strategies are totally different. So it is necessary to discuss MSR separately.
There are two types of high-risk nodes in MSR: (1) the event-related ones include all the nodes in the \emph{Perception} domain except \texttt{Preprocessing}. The robot uses those nodes to understand the conditions of the physical environment. (2) The action-related ones include all the nodes in the \emph{Control} domain which can directly interact with the actuator drivers. They are used to actively change the actual states of both the robot and environment. If either of these nodes are malicious, the robot and task can be compromised with unexpected consequences.

The rules to prevent MSR are determined by the missions and usage scenarios, which are usually specified by users. Table \ref{table:msrisks} lists some examples of MSRs and the corresponding rules in four scenarios. 
(1) In a domestic context, robots are designed to manage various human-centric tasks, e.g., house cleaning, baby-sitting.
They are required not to disturb human's normal life. 
(2) In a warehouse context, industrial robots are introduced to achieve high automation and improve the productivity, such as manipulator and autonomous ground vehicle (AGV).  
These robots are required to complete each subtask correctly, efficiently and safely.
(3) In a city context, autonomous vehicles and delivery robots move at high speeds in the transportation system, and handle complex events from outdoor dynamic environment. Thus, they need to obey the transportation rules and ensure the safety of passengers and public assets. 
(4) Robots are also deployed in many specialized scenarios to conduct professional missions. For example, rescue robots are used to search for survivors or extinguish fires. Medical robots are used in hospitals to diagnose and treat patients. Military robots are designed in battlefields to destroy enemies or constructions. These robots need to follow the rules related to their specific missions. 

\begin{table}[tb]
\caption{Examples of Mission-Specific Risks and Rules.}
\vspace{-10pt}
\centering
\resizebox{1\linewidth}{!}{
\begin{tabular}{l|l}
\hline
\rowcolor{gray!40} 
\textbf{Scenario}      & \multicolumn{1}{c|}{\textbf{Description}}   \\ \hline
\multirow{2}{*}{Domestic}  
			& \makecell[l]{The companion robot must send an alert when a user is in danger.} \\ 
			\cline{2-2} 
      & \makecell[l]{The robotic vacuum must be turned off when a user is sleeping.} \\ 
\hline
\multirow{2}{*}{\makecell[l]{Warehouse}}  
			& \makecell[l]{The manipulator must not grasp objects that exceed its limited weight.}   \\ 
			\cline{2-2} 
      & \makecell[l]{The AGV must recharge when the battery level is below a threshold.}   \\ 
\hline
\multirow{2}{*}{\makecell[l]{City}}          
			& The mobile vehicle must follow the traffic rule.   \\ 
			\cline{2-2} 
			& \makecell[l]{The mobile vehicle must maintain a safe distance with passengers.}   \\ 
\hline
\multirow{2}{*}{\makecell[l]{Specialized}} 
			& \makecell[l]{The firefighter robot must send an alert when detecting the wounded.}  \\ 
			\cline{2-2} 
      & \makecell[l]{The precision of the surgery robot must be above a specified threshold.}\\ 
\hline
\end{tabular}}
\label{table:msrisks}
\vspace{-15pt}
\end{table}

\subsection{Summary of Risks from Each Domain}
\label{sec:4-risk-verify}
An arbitrary malicious node in the robot app can incur the above risks. We discuss the potential risks and consequences caused by malicious functions in each domain.

\noindent{\textbf{Perception.}} 
If a node in the \emph{Perception} domain is untrusted, the robot states will be estimated as wrong values. Following the direct interactions, the robot will take anomalous actions, which violate the rules of MSR. Moreover, since the \texttt{Recognition} function typically adopts sensor fusion to reduce uncertainty caused by the physical limit of different sensors, such threat can cause GR as well.

For instance, an autonomous vehicle is navigating in a highway. A malicious \texttt{Preprocessing} function intentionally sends wrong sensory data to the \texttt{Object Recognition} function to cause optical illusions, e.g., recognizing a turn right sign as a stop sign. This will violate the traffic rule: ``vehicles cannot stop in a highway''.

\begin{table*}[tb]
\caption{Summary of risks, threats and mitigation for function interactions.}
\vspace{-10pt}
\centering
\resizebox{1.00\linewidth}{!}{
\begin{tabular}{|c|c|c|c|lll|} 
\hline
\rowcolor{gray!40}  \textbf{Risk} & \textbf{Domain Threat} & \textbf{Coordination Node} & \textbf{Executor} & \textbf{Policy} & \textbf{Parameter} & \textbf{Description} \\ 
\hline
\multirow{4}{*}{GR} & \multirow{4}{*}{\makecell[c]{Perception,\\Planning, Control}} & \multirow{4}{*}{GRCN} & \multirow{4}{*}{Developer} & Block & Block Bit & Allow/block the action of chosen flow. \\ 
\cline{5-7}
 & & & & FIFO\_Queue & Timeout	& Choose the action based on fifo order with time limit.	\\ 
\cline{5-7}
 & & & & Priority\_Queue & Timeout, Priority & Choose the action based on priority order with time limit. \\ 
\cline{5-7}
 & & & & Preemption & Priority & Choose the action based on priority order. \\ 
\hline
\multirow{3}{*}{RSR} & \multirow{3}{*}{Control}  & \multirow{3}{*}{RSRCN} & \multirow{2}{*}{Developer} & Block & Block Bit & Allow/block the velocity control action of chosen flow. \\ 
\cline{5-7}
 & & & & Safe & Threshold, Priority & Adjust max velocity based on fps data. \\ 
\cline{4-7}
 & & & End User & Constrain & Max\_vel\_limit & Limit adjustable max velocity limit with a user-defined value.  \\ 
\hline
MSR & \makecell[c]{Perception,\\Planning, Control} & MSRCN & End User & Block & Block Bit & Allow/block the action of chosen flow. \\ 
\hline
\end{tabular}}
\label{table:policy}
\vspace{-5pt}
\end{table*}

\noindent{\textbf{Planning.}} 
A malicious node in the \emph{Planning} domain can interrupt the current task, or reset the robot states to wrong values. In a common robot app, there can be multiple \texttt{Global Planner} functions for different goals based on various events from the \texttt{Recognition} functions\footnote{We merge all these interactions in our interaction graph (Figure \ref{fig:3-graph})}. This gives the malicious node chances to win the competition against other goals and compromise the robot states (GR). Besides, the malicious node can also directly modify the goal to make the robot take anomalous actions in a specific event (MSR).

For instance, a robot vacuum is executing the cleaning task in a living room. The \texttt{Global Planner} function is compromised and controlled by an adversary to set a new destination goal as the master bedroom for stealing privacy. This can violate a possible MSR rule: ``the robot vacuum cannot enter the bedroom''. If the robot does not have enough power to clean the master bedroom, this will violate the MSR rule: ``the AGV must recharge when the battery level is below a specified threshold.'' (Table \ref{table:msrisks}).

\noindent{\textbf{Control.}} 
If a function in the \emph{Control} domain is malicious, the adversary can launch attacks in three ways. First, the function can interrupt or suspend other actions from different interactions (GR). Second, it can increase the velocity to cause failures of image-related recognition functions through the indirect interaction (RSR). Third, it can directly control the robot to take unexpected actions in a specific scenario (MSR).

For instance, in a task of searching dangerous goods or wounded persons, the robot device receives images through the equipped camera at a certain frame rate. If the max\_vel node is malicious and intentionally increases the maximal velocity, there will be no or less correlation between adjacent frames. The \texttt{Image Recognition} function may fail to process each frame promptly, and frames containing safety-related information (e.g. drug, thief) can be missed. 

%% file: Tex/5-Methodology.tex
\section{Mitigation Methodology}
\label{sec:5-overview}
We present a novel methodology to mitigate the malicious function interactions. The core of our solution is a set of \emph{coordination nodes} (\cref{sec:5-2-deploy}) and \emph{security policies} (\cref{sec:5-3-policy}), as summarized in Table \ref{table:policy}.



\subsection{Coordination Node}
\label{sec:5-2-deploy}

\begin{figure}[tb]
\centering
\includegraphics[width=0.9\linewidth]{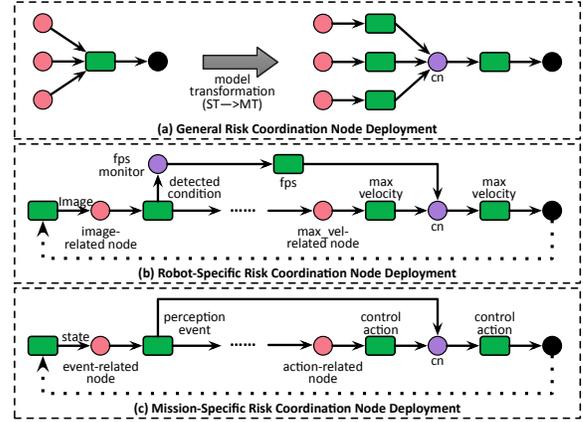}
\vspace{-10pt}
\caption{Three types of coordination nodes (purple circles). 
}
\label{fig:6-cn-deploy}
\vspace{-15pt}
\end{figure}

The coordination nodes are deployed inside the robot apps to regulate the interactions and enforce the desired security policies. They are designed to be general for different types of robots, function nodes and risks. Developers can deploy them into apps without modifying the internal function code. Users can adjust configurations based on their demands. We design three types of coordination nodes, to mitigate three types of risks respectively (Figure \ref{fig:6-cn-deploy}).

\noindent
\textbf{General Risk Coordination Node (GRCN).}
This node is inserted between the high-risk nodes and their successor node (Figure \ref{fig:6-cn-deploy}a). The published topics of each high-risk node need to be remapped to the subscribed topic of this GRCN to create new data flows, and the published topic of the GRCN need to be mapped to the subscribed topic of the successor node. Thus, the GRCN can control each data flow from the high-risk nodes based on various policies. 

\noindent
\textbf{Robot-Specific Risk Coordination Node (RSRCN).}
This node needs to coordinate the conflict between the image-related node and max\_vel-related node (Figure \ref{fig:6-cn-deploy}b). We use the same method to insert the RSRCN between the max\_vel-related node and its successor node. To collect the frame rate from the image-related node, we insert a fps\_monitor node to subscribe to the detected condition topic published by the image-related node. This fps\_monitor node measures the frequency of the triggered event and publishes the frame rate to the fps topic. The RSRCN subscribes to this fps topic and uses it as reference for max velocity adjustment. 

\noindent
\textbf{Mission-Specific Risk Coordination Node (MSRCN).}
This node needs to allow/block the actions taken under wrong conditions (Figure \ref{fig:6-cn-deploy}c). Thus, it is deployed between each action-related node and its successor, and subscribes to all perception event topics of event-related nodes. In this way, the MSRCN can collect all perception events in the app and obtain the control of each action. It is worth noting that there can be multiple GRCNs for each interaction, but the numbers of both RSRCN and MSRCN are always one. 

\subsection{Security Policies}
\label{sec:5-3-policy}
To mitigate the malicious interactions in an app, each type of coordination nodes implements a set of policies. 
Table \ref{table:policy} lists the policies we have built along with the descriptions and parameters for GRCN, RSRCN and MSRCN. Each policy needs to be configured by either the developer or end user, as shown in the ``Executor'' column.

\noindent
\textbf{GRCN Policies.} 
GRCN aims to coordinate data flows from different high-risk nodes. We use four types of policies to adapt to different scenarios. Specifically, the block policy is used when the user wants to stop the current action immediately in case of emergency. When multiple high-risk nodes publish control commands, the preemption policy will choose the action with the highest priority. For example, both the \texttt{Safe Control} and \texttt{Path Tracking} nodes publish velocity to the \texttt{Mobile Driver} node. However, the safe control action should be taken first because it is responsible for ensuring user's safety. FIFO\_Queue and Priority\_Queue policies are used for high-risk nodes with high requirements of completion time, such as search, rescue and obstacle avoidance. 

\noindent
\textbf{RSRCN Policies.} 
RSRCN aims to resolve the conflicts between data flows from the image-related (\emph{iflow}) and max\_vel-relate (\emph{vflow}) nodes. We use three types of policies to adjust the maximal velocity of the robot. Block policy allows/blocks the action from \emph{vflow} and does not affect the action from \emph{iflow}. Safe policy uses thresholds to bridge the maximal velocity with fps. Based on the fact that a higher velocity requires a faster processing capability, we assume the maximal velocity is proportional to the fps. Then the threshold serves as a scale factor and can be configured by users. Constrain policy sets a maximal velocity limit to ensure safety in complex and dynamic environments. This is particularly useful when users want the robots to work at low speeds psychologically even though they drive within safe speed ranges.

\noindent
\textbf{MSRCN Policies.} 
MSRCN aims to coordinate the conflicts between the data flows from the event-related node (\emph{eflow}) and action-relate node (\emph{aflow}). We only adopt block policy to decide whether the action should be taken under some specific conditions. However, the block bits of \emph{eflow} and \emph{aflow} are different. Bit 0/1 in \emph{aflow} denotes that the actions are allowed/blocked, while Bit 0/1 in \emph{eflow} represents whether the condition event is triggered or not. Thus, end users can control all the actions under arbitrary conditions.

To reduce the complexity of configuring our methodology for unexperienced end users, we delegate
part of the policy selection and parameter configuration tasks to the developers. It is reasonable because some risks are derived from the race condition while the others are caused by falling short of user's expectation. Specifically, the developers enforce appropriate policies for each GRCN and set the corresponding parameters. Moreover, the developers also preset the parameters in the block and safe policies for RSRCN based on the robot's characteristics. On the other hand, the end users only have the control of policy selection in RSRCN and MSRCN. The parameters they need to configure are just max\_vel\_limit in RSRCN and block bit in MSRCN. Table \ref{table:policy} shows the role of end users and developers for each policy (the ``Executor'' column).


%% file: Tex/6-RTron.tex
\section{System Design}
\label{sec:6-sys-design}
\begin{figure}[tb]
\centering
\includegraphics[width=0.99\linewidth]{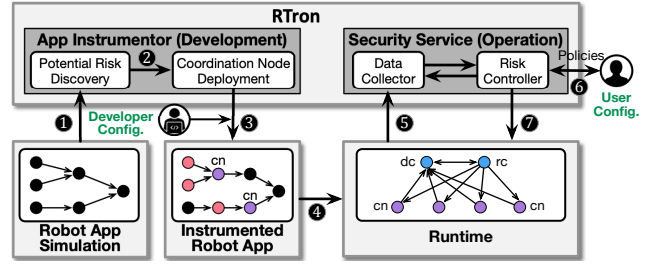}
\vspace{-5pt}
\caption{\textsc{RTron} system overview. }
\label{fig:5-rtron-overview}
\vspace{-20pt}
\end{figure}

We design \name, a novel end-to-end system equipped with the above mitigation. Given a potential vulnerable robot app, the developer first utilizes \name to add necessary coordination nodes to the app without modifying the original function node, and set up some security policies. Then the end user can safely launch the patched app on the robot, and configure other policies before the task starts. Figure \ref{fig:5-rtron-overview} gives the overview of \name. It consists of two components: (1) an \emph{App Instrumentor} for developers to detect potential risks in robot apps and deploys coordination nodes (\cref{sec:6-2-ins}); (2) a \emph{Security Service} that visualizes and configures the coordination nodes to mitigate risks at runtime (\cref{sec:6-3-ss}). 

\begin{algorithm}[bt]
\setstretch{0.9}
\caption{Algorithm for Potential Risk Discovery}
\footnotesize
\label{alg:6-risk-discovery}
\KwIn{$N$       \Comment{A set of nodes in a robot app}\\
\qquad\quad$T$    \Comment{A set of topics in a robot app}\\
\qquad\quad$N^p_j$  \Comment{A set of nodes publish to the topic $j$}\\
\qquad\quad$T^s_i$  \Comment{A set of topics subscribed by the node $i$}\\
\qquad\quad$T^p_i$  \Comment{A set of topics published by the node $i$}}
\KwOut{$RN$ \Comment{Risk nodes in a robot app}}
\ForEach{topics $t_j \in T$}
{
  \If{num($N^p_j$) \textgreater 1}
  {
    $RN^{ST}_{gr} \leftarrow \left\{N^p_j\right\}$\;
  }
  \If{(`max\_vel' $\in$ $t_j.name$) $\wedge$ ($t_j.type$ == `std\_msgs/Float64')}
  {
    $RN^{max}_{rsr} \leftarrow \left\{N^p_j\right\}$\;
  }
  \ForEach{string $s_n \in$ EVENT\_MSG\_TYPE}
  {
    \If{($s_n$ $\in$ $t_j.type$) $\vee$ (`detect' $\in$ $t_j.name$)}
    {
      $RN^{event}_{msr} \leftarrow \left\{N^p_j\right\}$\;
    }
  }
  \ForEach{string $s_n \in$ ACTION\_MSG\_TYPE}
  {
    \If{$s_n$ $\in$ $t_j.type$ $\vee$ (`goal' $\in$ $t_j.name$)}
    {
      $RN^{action}_{msr} \leftarrow \left\{N^p_j\right\}$\;
    }
  }
}
\ForEach{node $n_i \in N$}
{
  sort node's subscriptions $T^s_i$ by $T^s_i.type$\;
  \ForEach{subscription $s_k \in T^s_i$}
  {
    \If{$s_k.type$ == $s_{k+1}.type$}
    {
      $RN^{MT}_{gr} \leftarrow \left\{n_i\right\}$\;
    }
  }
  \ForEach{subscription $s_k \in T^s_i$}
  {
    \If{$s_k.type ==$ `sensor\_msgs/Image'}
    {
      \ForEach{publication $p_m \in T^p_i$} 
      {
        \ForEach{string $s_n \in$ RECOG\_TOPIC\_NAME}
        {
          \If{$s_n$ $\in$ $p_m.name$}
          {
            $RN^{image}_{rsr} \leftarrow \left\{n_i\right\}$\;
          }
        }
      }
    }
  }
}
\vspace{-5pt}
\end{algorithm}

\subsection{App Instrumentor}
\label{sec:6-2-ins}

The goal of this module is to instrument the target app's source code to make it compatible with \name. It patches an app with certain coordination nodes to collect events and actions from high-risk function nodes, and guard the robot at runtime. Two subcomponents are introduced to identify high-risk function nodes, and the locations to deploy the coordination nodes, respectively. 

\noindent\textbf{Potential Risk Discovery.}
This submodule is designed to help developers identify high-risk function nodes in a robot app. It first simulates the lifecycle of the target app and automatically generates the interaction graph offline. Then it traverses all function nodes (black circles in Figure \ref{fig:5-rtron-overview}) in the graph and identifies three types of high-risk function nodes: GR node $RN_{gr}$, RSR node $RN_{rsr}$ and MSR node $RN_{msr}$ (\ding{182}). 
Algorithm \ref{alg:6-risk-discovery} describes our identification strategy. 
We conclude one rule to discover each type of risky nodes:

\emph{GR Rule:}
we identify the topics in the graph whose indegree is greater than 1. All nodes that publish to these identified topics are denoted as $RN^{st}_{gr}$ with single topic (Lines 1-3). The node with more than one subscribed topics of the same message type can be integrated to $RN^{mt}_{gr}$ with multiple topics (Lines 12-16).

\emph{RSR Rule:}
to identify the image-related node $RN^{image}_{rsr}$ and max\_\\vel-related node $RN^{max}_{rsr}$, \name checks the topic name and type of each subscribed or published message (Lines 4-5,17-22). It searches the key words (e.g., `detect', `people' and `face') in the RECOG\_TOPIC\\\_NAME string list. Evaluations in \cref{sec:7-evaluation} indicate this key word searching can effectively identify the RSR nodes. 

\emph{MSR Rule:}
to identify the event-related node $RN^{event}_{msr}$ and action-related node $RN^{action}_{msr}$, \name checks if the message type of each topic (Lines 6-11) is in the EVENT\_MSG\_TYPE or ACTION\_MSG\_\\TYPE lists since message types typically use standard ROS naming conventions \cite{ros-msg}. The complete lists of EVENT\_MSG\_TYPE and ACTION\_MSG\_TYPE are presented in Appendix \ref{sec:A-msg-type}.

\noindent\textbf{Coordination Node Deployment.}
The collected information of potential risks is used to configure the coordination node setting(\ding{183}). This includes a set of topics and parameters. Topics represent the state transition between two function nodes: the subscribed and published topics specify the predecessor and successor nodes of each coordination, respectively. The parameters are used to expose an interface to the end user for configuring each policy. With these configuration files, a \emph{Coordination Node Deployment} submodule is designed to deploy coordination nodes into the app automatically (\ding{184}). 
Meanwhile, the developers check the details of the risks, select the optional policies for GRCN and configure related parameters.

Taking GRCN as an example (Figure \ref{fig:6-console}(a) in Appendix \ref{sec:B-user-console}). GRCN monitors velocity data from three risky nodes: \texttt{Navigation Control}, \texttt{Tele-operation} and \texttt{Safe Control}. The data transmission of each node is marked as flow1, flow2 and flow3. The developer can select the Priority\_Queue policy after the app is launched, and set flow3 from \texttt{Safe Control} as the highest priority, indicating its velocity action should be always taken first. However, if the coordination node cannot receive the responding actions before the user-defined timeout (i.e. 0.2s), it will transmit the velocity action of flow2 with the second priority. 

\subsection{Security Service}
\label{sec:6-3-ss}
\begin{figure}[tb]
\centering
\includegraphics[width=0.8\linewidth]{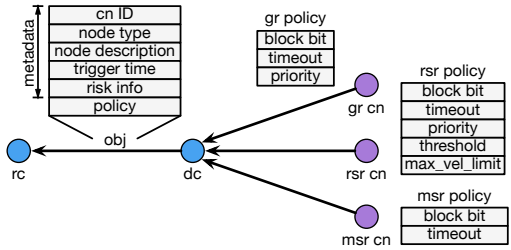}
\vspace{-10pt}
\caption{Risk model of three types of risks in \emph{Data Collector}.}
\label{fig:6-risk-model}
\vspace{-15pt}
\end{figure}

This module aims at visualizing and mitigating risks of malicious interactions at runtime. It consists of two subcomponents deployed along with the robot app. 

\noindent\textbf{Data Collector.}
When the robot executes the app within the environment, all coordination nodes in the instrumented app keep forwarding their information to this submodule (\ding{185}). Such information is stored as a risk model, which consists of metadata and a set of policy parameters (\ding{186}). As shown in Figure \ref{fig:6-risk-model}, the metadata records basic information of a coordination node, including its ID, node type, node description, trigger time and risk information. They manage each coordination node and visualize to the users for risk display and policy configuration.

\noindent\textbf{Risk Controller.}
This submodule visualizes risk information and enforces policies from users to each coordination node. Right after the app is launched on the robot, the \emph{Risk Controller} obtains all the information of each coordination node from the \emph{Data Collector}. It then configures each coordination node by sending user-defined policy parameters (\ding{188}). When the \emph{Data Collector} receives an event and the corresponding coordination nodes' actions at runtime, the \emph{Risk Controller} evaluates them against a collection of security policies. Some policies are mandatory, while some are optional, depending on the real-world demands (e.g. task or scenario) of end users. 

The \emph{Risk Controller} provides an interface for end users to check the details of risks and select the optional policies (\ding{187}). Appendix \ref{sec:B-user-console} presents the user consoles for three types of coordination nodes. 
The console displays a visual representation of the rule violation and available policies based on the risk model of each node. Specifically, the node type, node description, trigger time and risk information are summarized as the violated rule, violation cause, violation details respectively. 
Note that the end users only have full control of policy selection for RSRCN and MSRCN, and parameter configurations for two specific policies.

Taking RSRCN as an example (Figure \ref{fig:6-console}(b) in Appendix \ref{sec:B-user-console}). End users can check the current violation information and reset the corresponding policy parameters at runtime. When a robot moves from an obstacle-free environment (e.g., Highway) to a complex environment (e.g. downtown area), users can select the Constrain policy in an RSRCN to limit the robot's maximal velocity.

%% file: Tex/7-Evaluation.tex
\section{Evaluation}
\label{sec:7-evaluation}
\begin{figure*}[tb]
\centering
\includegraphics[width=1.0\linewidth]{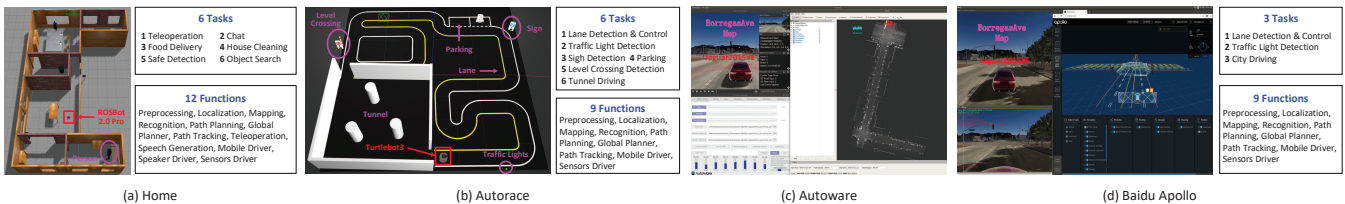}
\vspace{-22pt}
\caption{Four simulated scenarios in the Gazebo/LGSVL.}
\label{fig:7-gazebo-model}
\vspace{-10pt}
\end{figure*}

We aim to answer the following questions:

\begin{packeditemize}
	\item Can \textsc{RTron} effectively detect three types of interaction risks? 
	What is the relationship between the interaction risks and task characteristics in each robot app? (\cref{sec:7-1-discovery})
	\item How many coordination nodes are required to deploy in a typical robot app? How to configure the policy for an end user under various environmental contexts? (\cref{sec:7-2-cn})
	\item What is the performance overhead of \textsc{RTron}? (\cref{sec:7-3-overhead})
	\item How practical is it to launch end-to-end attacks and perform risk mitigation on physical robots and environments? (\cref{sec:7-4-usecase})
\end{packeditemize}

\noindent
\textbf{Target Workloads.} 
In addition to a total of 110 open-source single-functional apps from the ROS showcase website \cite{ros-robot}, we perform analysis of more complex apps (Figure \ref{fig:7-gazebo-model}):

\begin{packeditemize}
	\item \emph{Home scenario}: home-based apps and robots are used to accompany people and conduct housework. These tasks include teleoperation, chat, food/drink delivery, cleaning, safe detection, and object search. We use four ROS apps (Remote Control, Face/Person Detection, Object Search and Voice Interaction) of RosBot 2.0 Pro \cite{rosbot} to develop one home app (Figure \ref{fig:7-gazebo-model}a). 

	\item \emph{AutoRace scenario \cite{app-autorace}}: this type of apps is designed for competition of autonomous driving robot platforms. To ensure that the robot can drive on the track safely, there are six necessary missions for the robot to execute, including lane detection \& control, traffic light detection, sign detection, parking, level crossing detection and tunnel driving. 
	We use the open-source Autonomous Driving app of Turtlebot3 \cite{turtlebot3} which can realize all six tasks in the autorace scenario (Figure \ref{fig:7-gazebo-model}b).

	\item \emph{Autonomous driving scenario}: we consider two mainstream self-driving apps: Autoware \cite{app-autoware} and Apollo \cite{app-apollo}, which have been fully deployed and tested in physical autonomous vehicles. These two apps are more complex than the AutoRace scenario, with a richer set of self-driving modules composed of sensing, computing, and actuation capabilities (Figure \ref{fig:7-gazebo-model}c and \ref{fig:7-gazebo-model}d).
\end{packeditemize}

\noindent
\textbf{Experimental Setup.} 
Since this paper focuses on the software risks in robot apps, 
we mainly use simulation to validate our solution. Implementation and evaluation on physical robots will be demonstrated in \cref{sec:7-4-usecase}.
We choose the Gazebo simulator \cite{gazebo} and ROS Kinetic in the home and autorace scenarios, which run on a server equipped with 1.6GHz 4-core Intel i5 processor and Nvidia MX110 GPU. In the autonomous driving scenario, we use the LGSVL simulator \cite{lgsvl} with ROS Indigo for Apollo 3.5, and ROS Melodic for Autoware 1.14, running on a server with 4.2GHz 8-core Intel i7 and Nvidia GTX 1080 GPU. 
We use Rviz \cite{rviz} to visualize 3D information from both the simulator and robot apps. 


\subsection{Risk Identification}
\label{sec:7-1-discovery}

\noindent\textbf{Single-functional Apps.} 
We successfully extract all the GRs and MSRs from all 110 open-source apps. GRs are identified by checking the nodes and topics based on their topology relationship. Some GRs are ignored  when they publish messages to the log/visualization topic, which will not bring risks to the robot app. MSRs are identified by inspecting if the standardized topic types are matched. 

Different from the GR rule, the RSR rule involves the identification of specific topic names and types. We choose 15 image-related apps (e.g., Face/Person Detection, Object/Scene Detection, Object Search, Autonomous Driving) and 1 max\_vel-related app (Autonomous Driving). We successfully discover all 20 image-related and 4 max\_vel-related RSRs from these apps. 




\begin{figure}[tb]
\centering
\includegraphics[width=1.00\linewidth]{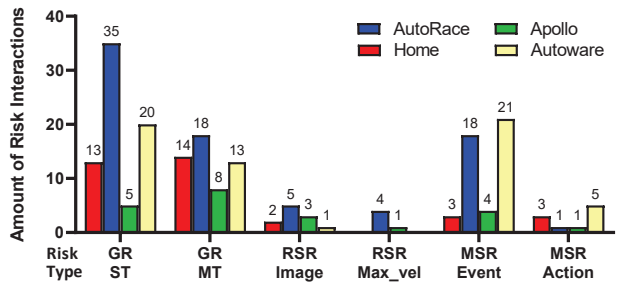}
\vspace{-20pt}
\caption{Numbers of high-risk nodes in four robot apps.}
\label{fig:7-risk-stat}
\vspace{-18pt}
\end{figure}

\noindent\textbf{Multi-functional Apps.} 
\name is also scalable for analysis of more complex apps. As examples, we show the interaction graph of the home-based app produced by \name in Figure \ref{fig:C-home-rosgraph} , and the high-risk node for each type in the home and autorace apps in Table \ref{table:risk_examples} in Appendix \ref{sec:C-app-graph}.

\name successfully identifies 198 risk interactions in the four target apps. Figure \ref{fig:7-risk-stat} lists the numbers of extracted nodes with respect to each risk type. We can observe the numbers of risk interactions in the autorace (blue bar) and autoware (yellow bar) apps are larger than home (red bar) and apollo (green bar) apps, although the home app has the largest number of functions. This is caused by the differences in the internal structure of each robot app. In the home scenario, each task is relatively independent. 
However, in the autorace and autoware apps, all tasks are organized as a monolithic component to control the robot to drive safely. To achieve this, these two apps need to recognize various scenes from sensory images and take the corresponding actions. Consequently, the high dependency among those tasks increases the number of GRs. Moreover, the requirement of image and scene recognition increases the number of image-related RSRs and event-related MSRs. Apollo is a special case where the number of topics is far smaller than the other apps, thus the number of risks is also the smallest (Table \ref{table:tool_overhead}).


\subsection{Risk Mitigation}
\label{sec:7-2-cn}

\noindent\textbf{CN Analysis.} 
\name uses the extracted risk information to deploy CNs. For GRs, the number of GRCNs depends on the number of high-risk interactions linked to the same node. Thus, \name checks the GR information of ``Pub Node'' and deploys the GRCN between high-risk nodes and their pub node. For RSRs, since RSRCNs directly publish velocity messages to the \texttt{Mobile Driver} function, the number of RSRCN is always 1. The subscriptions of RSRCN is related to the number of image-related nodes and max\_vel-related nodes. Besides, as described in \cref{sec:5-2-deploy}, each image-related node should be assigned to an fps\_monitor node to generate the 
processing rate of the image recognition process. So the number of required fps\_monitor nodes depends on the number of image-related nodes. For MSRs, the number of MSRCNs is 1, as all event-related and action-related nodes publish corresponding messages to the MSRCN, which then sends the action message to all related actuator driver nodes.

Table \ref{table:cn_stats} lists the numbers of three types of CNs in the four robot apps. We observe that GRCNs account for a large portion of the total added nodes. Due to a large number of RSR image-related interactions, the autorace app has more fps\_monitor nodes than the home app. The complete interaction graph with added CNs is shown in Figure \ref{fig:C-home-cn-rosgraph}, Appendix \ref{sec:C-app-graph} for the example of the home app.

\begin{table}[tb]
\caption{Numbers of CNs in four complex robot apps.}
\vspace{-10pt}
\centering
\arrayrulecolor{black}
\resizebox{0.95\linewidth}{!}{
\begin{tabular}{|c|c|c|c|c|c|c|} 
\hline
\rowcolor{gray!40} {\cellcolor{gray!40}} & \multicolumn{3}{c|}{\begin{tabular}[c]{@{}>{\cellcolor{gray!40}}c@{}}\textbf{GRCN}\\ \end{tabular}} & \multicolumn{2}{c|}{\textbf{RSR}} & {\cellcolor{gray!40}} \\ 
\hhline{|>{\arrayrulecolor{gray!40}}->{\arrayrulecolor{black}}----->{\arrayrulecolor{gray!40}}->{\arrayrulecolor{black}}|}
\rowcolor{gray!40} \multirow{-2}{*}{{\cellcolor{gray!40}}\textbf{Scenario}} & \textbf{Perception} & \textbf{Planning} & \textbf{Control} & \textbf{FMN} & \textbf{CN} & \multirow{-2}{*}{{\cellcolor{gray!40}}\begin{tabular}[c]{@{}>{\cellcolor{gray!40}}c@{}}\textbf{MSR}\\\textbf{CN}\\ \end{tabular}}  \\ 
\hline
Home & 8 & 3 & 1 & 2 & 1 & 1 \\ 
\hline
AutoRace & 16 & 2 & 4 & 5 & 1 & 1 \\
\hline
Apollo & 4 & 1 & 1 & 3 & 1 & 1 \\
\hline
Autoware & 11 & 3 & 2 & 1 & 1 & 1 \\
\hline
\end{tabular}}
\label{table:cn_stats}
\vspace{-5pt}
\end{table}

\begin{table}
\caption{High-risk interacted topics and features of three GRCN types in the home app.}
\vspace{-10pt}
\centering
\resizebox{0.95\linewidth}{!}{
\begin{tabular}{|c|l|c|} 
\hline
\rowcolor{gray!40}
\textbf{CN Type} & \multicolumn{1}{c|}{\textbf{Interacted Topics}} & \textbf{Feature} \\ 
\hline
Perception & \begin{tabular}[c]{@{}l@{}}`/explore\_server/status', `/move\_base/status', \\`tf', `tf\_static', `/camera/rgb/image\_raw', \\`/camera/depth/image\_raw', \\`/move\_base/global\_costmap/footprint', \\`/move\_base/local\_costmap/footprint'\end{tabular} & \begin{tabular}[c]{@{}c@{}}State\\Parallelization\end{tabular}  \\ 
\hline
Planning & \begin{tabular}[c]{@{}l@{}}`/move\_base/goal', `/move\_base/cancel', \\`/move\_base\_simple/goal'\end{tabular} & \begin{tabular}[c]{@{}c@{}}Goal\\Queuing\end{tabular} \\ 
\hline
Control & `/cmd\_vel' & \begin{tabular}[c]{@{}c@{}}Action\\Preemption\end{tabular} \\
\hline
\end{tabular}}
\label{table:grcn_type}
\vspace{-10pt}
\end{table}

\noindent\textbf{Policy Selection.} 
\textsc{RTron} implements a variety of policies for three types of CNs. How to select the appropriate policy for each CN is critical for the secure operation of robot apps. We use the home app as an example to illustrate the guideline for policy selection. 

\emph{GRCN}: 
this is designed to coordinate direct high-risk interactions between multiple connected nodes. Based on the types of interacted topics, we classify GRCN into three categories: perception, planning and control. As shown in Table \ref{table:grcn_type}, the messages of interacted topics in perception are related to the sensory information (e.g. images) or preprocessed robot states (e.g. footprints, status). Typically, multiple messages with the same type are published to the same target node, and processed in parallel for either sensor fusion or state monitoring. Thus, there is no contention among these messages. 

Messages of the interacted topics in planning or control contend with each other to get the long-term and instant control of the robot. Specifically, when a message of a new planning goal is received, the robot must first complete the previous goal before executing the current one. For example, an object search task is launched after the \texttt{search\_manager} node publishes a goal to the \texttt{/move\_base\_simple/goal} topic. An adversary can use a malicious \texttt{rviz} node to send another arbitrary destination to this topic (Figure \ref{fig:C-home-rosgraph} in the appendix). The object search task will be immediately interrupted and then the robot is controlled to reach the designated position. Thus, a GRCN with the `FIFO\_Queue' or `Priority\_Queue' policy can delay such malicious actions without task interruption. 

Different from the planning messages, the control messages need to control the robot immediately. End users can select the `Preemption' policy of GRCN for coordination. For instance, the malicious \texttt{teleop\_twist\_keyboard} node can flood the \texttt{/cmd\_vel} topic while the robot is following a planned path to the destination. Then the topic receives the messages from both \texttt{teleop\_twist\_keyboard} and \texttt{move\_base} nodes simultaneously, which causes the robot to switch velocity in the two target directions. 
By assigning the highest priority to the move\_base-related velocity control interaction (i.e. \texttt{/cmd\_vel}), the \texttt{move\_base} node can control the robot first. 

\begin{table}
\caption{Processing time of potential risk discovery.}
\vspace{-10pt}
\centering
\arrayrulecolor{black}
\resizebox{0.95\linewidth}{!}{
\begin{tabular}{|l|c|c|c|c|c|} 
\hline
\rowcolor{gray!40} {\cellcolor{gray!40}} & {\cellcolor{gray!40}} & {\cellcolor{gray!40}} & \multicolumn{3}{c|}{\textbf{Processing Time (s)}}  \\ 
\hhline{|>{\arrayrulecolor{gray!40}}--->{\arrayrulecolor{black}}---|}
\rowcolor{gray!40} \multirow{-2}{*}{{\cellcolor{gray!40}}\textbf{Application} } & \multirow{-2}{*}{{\cellcolor{gray!40}}\begin{tabular}[c]{@{}>{\cellcolor{gray!40}}c@{}}\textbf{Node} \\\textbf{Number}\end{tabular}} & \multirow{-2}{*}{{\cellcolor{gray!40}}\begin{tabular}[c]{@{}>{\cellcolor{gray!40}}c@{}}\textbf{Topic}\\\textbf{Number}\end{tabular}} & \textbf{GR} & \textbf{RSR} & \textbf{MSR}  \\ 
\hline
Teleoperation \cite{app-teleop}        	& 4  & 17  & 0.114 & 0.113 & 0.057 \\ 
\hline
Voice Interaction \cite{app-voice}    	& 6  & 7   & 0.035 & 0.035 & 0.011 \\ 
\hline
Mapping \cite{app-mapping}              & 6  & 25  & 0.308 & 0.299 & 0.152 \\ 
\hline
Navigation \cite{app-navi}           	& 8  & 63  & 0.764 & 0.727 & 0.498 \\ 
\hline
Exploration \cite{app-explore}          & 10 & 84  & 1.12  & 1.086 & 0.753 \\ 
\hline
Home             	 					& 21 & 125 & 3.121 & 3.199 & 1.927 \\ 
\hline
AutoRace \cite{app-autorace}            & 25 & 112 & 4.075 & 4.049 & 2.105 \\
\hline
Apollo \cite{app-apollo}				& 21 & 39  & 0.631 & 0.606 & 0.306 \\ 
\hline
Autoware \cite{app-autoware}            & 38 & 218 & 2.945 & 2.931 & 1.747 \\
\hline
\end{tabular}}
\label{table:tool_overhead}
\vspace{-15pt}
\end{table}

\emph{RSRCN:} end users are not recommended to set the `Block' or `Safe' policy. These two options should be chosen by app developers after extensive evaluations. Instead, users can choose the `Constrain' policy to set a maximal velocity value to limit the robot's speed. This is very effective and safe, especially when the robot's working environment is highly complex and dynamic, and the task completion time is not very critical. For example, if an adversary compromises the \texttt{move\_base} node and increases the robot's speed to a dangerous level, this can cause a potential traffic accident. By setting an appropriate threshold in the `Safe' policy or max\_vel\_limit in the `Constrain' policy, the robot will slow down its speed without object detection failures.

\emph{MSRCN:} although there is only one policy option, users can customize different rules to allow/block the actions of specific robots under specific conditions. Taking the home app as an example, the MSRCN receives messages from three event-related topics (\texttt{/objects}, \texttt{/person\_detector/detections}, \texttt{/odom}) and two action-related topics (\texttt{/audio/audio}, \texttt{/cmd\_vel}). Users can set a rule to disallow the robot's movement when it detects the target object. This can identify and mitigate the interruption of the object search task caused by the malicious \texttt{rviz} node mentioned above.


\begin{figure}[tb]
\centering
\includegraphics[width=0.8\linewidth]{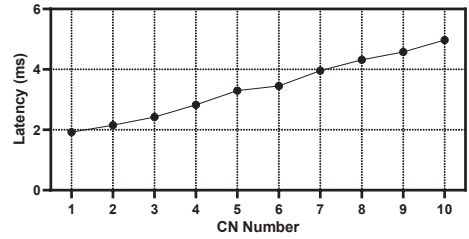}
\vspace{-10pt}
\caption{Overhead of CNs in an end-to-end data flow.}
\label{fig:7-cn-overhead}
\vspace{-15pt}
\end{figure}

\begin{figure*}[tb]
\centering
\includegraphics[width=0.95\linewidth]{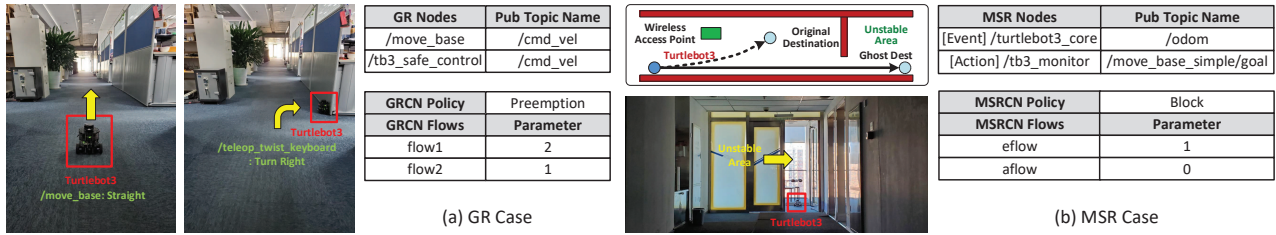}
\vspace{-8pt}
\caption{GR and MSR experiments on turtlebot3.}
\label{fig:7-usecases}
\vspace{-10pt}
\end{figure*}

\subsection{Performance Overhead}
\label{sec:7-3-overhead}

\noindent\textbf{Offline overhead.} 
We evaluate the risk discovery stage of \textsc{RTron} in terms of processing time for identifying high-risk nodes in a robot app.
Table \ref{table:tool_overhead} reports the performance results of 9 robot apps with different numbers of topics and nodes. We repeat each experiment for 20 times to calculate the average latency. We conclude that the risk discovery has negligible overhead as an offline process. The results also show that the processing time is affected by the number of topics and nodes. This is because the risk identification depends on the traversal of either nodes or topics (Algorithm \ref{alg:6-risk-discovery}). Specifically, there are two iterations in the process of both GR and RSR discovery and one iteration of topics in the process of MSR discovery. Thus, discovering GR takes similar time as RSR, which is longer than MSR. One exception is the autorace app, which has the largest processing time, but fewer nodes and topics than the home app. This is because there are more high-risk GR interactions in the autorace app (Table \ref{table:cn_stats}), which add extra work (i.e. related topic type and name match) in the node iteration process.

\noindent\textbf{Runtime overhead.} 
This includes the overhead from the coordination nodes and security service. The security service is only responsible for risk monitoring and policy configuration of each coordination node, without any interference on the execution of the robot app. Much like IoT policy enforcement systems \cite{iot-iotMon,iot-iotGuard}, we ignore the overhead of this process since users manually configure the policy for each CN only at the mission launch stage or scenario change condition. 
The coordination nodes are distributed among function nodes in the robot app, which can increase the end-to-end latency from the perception to the control stages. Although there are dozens of nodes in a typical robot app, these nodes work in a parallel multi-flow mode. To achieve real time, typically each data flow includes fewer than 10 nodes. So we consider the overhead of end-to-end latency within 10 coordination nodes. As shown in Figure \ref{fig:7-cn-overhead}, the extra latency incurred by 10 coordination nodes is around 5ms. This is trivial even for the autonomous driving app with the strongest real-time constraint: according to the industry standards published by Mobileye \cite{mobileye} and design specifications from Udacity \cite{udacity}, the latency for processing tragic condition in an autonomous driving app should be within 100 ms, which is far larger than the overhead of coordination nodes.

\subsection{Case Studies in the Real World}
\label{sec:7-4-usecase}
To demonstrate the practicality of the considered threats and proposed solution, we implement and evaluate several scenarios in a physical device, i.e., Turtlebot3. Figure \ref{fig:7-usecases} shows our settings and real-world environment. The Turtlebot3 is an open-source mobile base equipped with a Raspberry Pi CPU@1.3GHz, 1 GB memory and a 360 Laser Distance Sensor (LDS), running ubuntu 16.04 and ROS kinetic. It is connected to a server (Intel i7 CPU@4.2GHz with 16GB of RAM) for computation offloading and mission launching.

\noindent
\textbf{Attack Method.} 
We develop two normal tools \texttt{/tb3\_safe\_control} and \texttt{tb3\_monitor} to monitor and control the robot's movement. We insert some malicious codes in these tools which send wrong control commands. The detailed implementation of our attacks is described in Appendix \ref{sec:D-attack-impl}. We encapsulate these tools into two ROS packages and successfully upload them to the ROS platform as a developer\footnote{To avoid raising ethical concerns, we add an extra trigger such that the attacks happen only when the MAC address of the robot matches a predefined one. This ensures that the malicious package will not affect normal users. The repo is uploaded to the ROS platform from Dec. 5, 2020 to the date of writing on the website: link removed for anonymity, as it contains the author's information}. \emph{This validates our threat model that an adversary can easily share malicious packages in the ROS platform.} Next, we download these two packages as another developer, and implement them on the Turtlebot device. Below we describe the malicious behaviors and how our system can mitigate them with two cases\footnote{Considering the potential physical damages caused by vehicle's high speeds, the RSR case is implemented in the simulator in Appendix \ref{sec:D-rsr-exp}}.

\noindent
\textbf{GR Case.} 
The \texttt{/tb3\_safe\_control} node generates malicious velocity commands during the robot's navigation at certain moments. In Figure \ref{fig:7-usecases}a, the robot plans a straight route in the corridor. During its movement, the \texttt{/move\_base} node computes the real-time velocity and publishes it to the \texttt{/cmd\_vel} topic to drive the robot to the destination. Due to the shared state between these two nodes, the malicious node compromises the robot and creates a crash through publishing continuous ``turn right'' commands to the shared topic. In \textsc{RTron}, the developer can choose \emph{Preemption} policy in the GRCN and set different priorities to each flow. Then the malicious node is not able to interrupt the normal navigation behaviors in this case.


\noindent
\textbf{MSR Case.} 
The \texttt{tb3\_monitor} node sends malicious goal commands during the robot's navigation at certain moments. As shown in Figure \ref{fig:7-usecases}b, it generates a wrong destination in an unstable area far away from the wireless access point. As part of the computations is offloaded to the remote server, the robot running into this area will lose network connection, and malfunction. In our experiment, after the destination is changed, the robot navigates into the unstable area and finally stops under the poor network condition. In \textsc{RTron}, we choose \emph{Block} policy in the MSRCN and set different parameters to each flow. The developer uses position from the \texttt{/odom} topic to implement a function node to check whether the robot moves in the unstable area. This node is connected with MSRCN and marked as an eflow. In this way, any suspicious destination within the unstable area would be blocked.

%% file: Tex/8-RelatedWork.tex
\section{Related Works}
\label{sec:8-related-work}
\noindent
\textbf{Robotic Security.} 
Existing research on robotic security has mainly focused on traditional security issues in robot systems, e.g., network communication \cite{ros-authen-1,ros-authen-2,ros-authen-3}, denial-of-service attacks \cite{ros-vulnerable} and software vulnerabilities \cite{buf-overflow-1,buf-overflow-2, buf-overflow-3, sec-analysis,input-bugs}. In addition, adversaries can also spoof the sensory data (\cite{spoof-gps1, spoof-gps2, spoof-gps3, spoof-gps4, spoof-gps5,spoof-lidar-1, spoof-lidar-2,spoof-optical,spoof-gyroscopic-1, spoof-gyroscopic-2,spoof-gyroscopic-3,inject-magnetic}), fake the actuator signals \cite{spoof-signal}, or tamper with the micro-controller input \cite{sec-analysis}.



In this paper, we focus on a new type of security issue in robot apps, caused by malicious interactions. We are the first to demonstrate the feasibility and severity of this threat, as well as a possible defense solution against it. 


\noindent
\textbf{Interaction Risk Mitigation.}
Prior works studied the interaction risks in IoT apps \cite{iot-sift,iot-salus,iot-iotMon,iot-iotGuard,iot-soteria,iot-iostSan,iot-homeGuard,iot-menShen,iot-autoTap,iot-iRuler}. Users adopt operation rules following the “If-This-Then-That” (IFTTT) trigger-action programming paradigm \cite{ifttt-1,ifttt-2} to express automation behaviors among IoT devices. 
These methods translate the rules to the interaction graph, and verify if conflicts or policy violations can occur between interactions. 

There are three major differences between the interaction risks of robot apps and IoT apps. (1) For interaction modeling, robot apps not only inherit all the interactions from IoT apps, but also enjoy robot-specific ones, e.g., direct interactions via sharing internal states, indirect interactions caused by mobility. 
Robot apps need to cooperate with multiple functions, and require more complicated rules than the IFTTT model in IoT apps. (2) For risk identification, IoT apps are implemented by verifying if the interaction between different rules violates user-defined policies. However, robot apps have not only such risks (MSR), but also new ones (GR and RSR) due to data competition and mobility. (3) For risk mitigation, different from the simple ``allow/block'' policy adopted in IoT works, coordination of each type of risks needs a set of different configurable policies to mitigate malicious function interactions. All these distinct features of robot apps require new studies about the risk analysis and mitigation solutions, as we present in this paper.

%% file: Tex/9-Conclusion.tex
\section{Conclusion}
\label{sec:9-conclusion}

Function interaction provides great flexibility and convenience for robot app development. However, it also introduces potential risks that can threaten the safety of robot operations. This is exacerbated by the fact that current robot app stores do not provide security inspection over the function packages. We present the first study towards the safety issues caused by suspicious function interaction in robot apps. We introduce a novel end-to-end system and method to enforce security policies and protect the function interactions in robot apps. We hope this study can open a new direction for robotics security, and increase people's awareness about the importance of function interaction protection.



%% file: Tex/Appendix.tex
\clearpage


\section{Implementation of Code Analysis}
\label{sec:A-auto-analysis}
As discussed in the \cref{sec:3-1-model}, we require an understanding of how many function nodes in the interaction graph. 
In this section, we describe our approach to automatically map various repos in the ROS platform to several functions.

\subsection{Key Information Extraction}
To identify the function type of a ROS repo, three particular attributes of the repos can be inspected: 
\begin{enumerate}
  \item the \emph{repo name}, which can directly reflect the functionality of this repo; 
  \item the \emph{manifest file} (i.e. package.xml), which shows a functional brief of the repo. 
  \item the \emph{related document} (i.e. README file), which presents the detailed information of this package, including function description, topics and services; 
\end{enumerate}

\noindent
Listing \ref{listing-1} shows an example of the three attributes in rrt\_exploration repo. 
We regard the package name as one piece of key information and extract other two pieces of key information from the manifest file and related document. 
Specifically, we identify the text within the key tags (i.e. <description></description>) as key information in the manifest file. 
For the related document, we need first filter out the useless interference, such as the installation command and requirements introduction in the example. 
Then we use the remaining description of the repo as the key information.

\lstset{style=mystyle-1}
\begin{lstlisting}[language=Ant, caption=An example of the three attributes, label=listing-1]
`\textcolor{ForestGreen}{*** \textbf{the repo name} ***}`
rrt_exploration
`\textcolor{ForestGreen}{*** \textbf{the manifest file} ***}`
<description>A ROS package that implements a multi-robot 
RRT-based map exploration algorithm. It also has the 
image-based frontier detection that uses image processing 
to extract frontier points</description>
......
`\textcolor{ForestGreen}{*** \textbf{the related document} ***}`
It is a ROS package that implements a multi-robot map 
exploration algorithm for mobile robots. It is based on 
the Rapidly-Exploring Random Tree (RRT) algorithm. It 
uses occupancy girds as a map representation.The package 
has 5 different ROS nodes:
(1) Global RRT frontier point detector node.
(2) Local RRT frontier point detector node.
......
1. Requirements
The package has been tested on both ROS Kinetic and ROS 
Indigo, it should work on other distributions like Jade. 
`
\colorbox{CodeGray}{\$ sudo \ apt-get \ install \ ros-kinetic-gmapping}
`
......
2. Installation
Download the package and place it inside the /src folder 
in your workspace. And then compile using catkin_make.
......
\end{lstlisting}


\subsection{Function Classification}
To categorize 941 repos into 17 types of robotic functions in Table \ref{table:repository}, we first select 500 repos as a training set and use our expert experience to manually conclude the rules for each type from three pieces of key information in each repo. Then we use string regular expression to match these rules with tokens in the key information over above the rest of 441 repos. Finally, we manually analyze these 341 repos to verify the correctness of our automatic classification. Taking the \texttt{Visualization} function as an example, the rule is to determine whether the key information contains one of two kinds of tokens. One token is the function-related words and their variants, such as visualize and simulation. The other token is the function-related tool names, such rqt and gazebo. These tools are designed to visualize the robot and different tasks. Due to the fact that the more information probably hides more inferences, we identify the function types from repo name, manifest file and related documents successively, which means the matching process is finished if succeeds in one piece of key information. The rest of repos cannot be detected automatically will be analyzed manually.

Table \ref{table:automation} shows the results of our automated function classification. The automation rate is the radio of the successfully identified number to the sum of repos in this function. 
We can observer that the automation rate of most function types is more than 80\% except the \texttt{Support} and \texttt{Extension} functions. Fortunately, the repos in the \texttt{Visualization}, \texttt{Support} and \texttt{Extension} belong to the \emph{Others} domain, they are independent of the function nodes in the interaction graph. Thus, considering the repos in the first 14 function types, the automation rate of function node can reach 88.52\%. Among these successfully automated classified repos, the accuracy of our classification can achieve 99.82\%.

\begin{table}[tb]
\caption{The number of successful identifications in 941 repos.}
\vspace{-10pt}
\centering
\resizebox{1.0\linewidth}{!}{
\begin{tabular}{|l|c|c|c|c|c|}
\hline
\rowcolor{gray!40} 
\textbf{Function Type}                       & \textbf{\begin{tabular}[c]{@{}c@{}}Repo \\ Name\end{tabular}} & \textbf{\begin{tabular}[c]{@{}c@{}}Manifest\\ File\end{tabular}} & \textbf{\begin{tabular}[c]{@{}c@{}}ReadMe\\ File\end{tabular}} & \textbf{Manual} & \textbf{\begin{tabular}[c]{@{}c@{}}Automation\\ Rate\end{tabular}} \\ \hline
Preprocessing                                & 18                & 41                & 13           & 12              & 85.71\%                  \\ \hline
Localization                                 & 17                & 13                & 3            & 2               & 94.29\%                  \\ \hline
Mapping                                      & 15                & 11                & 4            & 1               & 96.77\%                  \\ \hline
Recognition                                  & 21                & 23                & 4            & 2               & 96.00\%                  \\ \hline
Path Planning                                & 44                & 49                & 5            & 5               & 95.15\%                  \\ \hline
Goal Planner                                 & 5                 & 4                 & 0            & 2               & 81.82\%                  \\ \hline
Path Tracking                                & 8                 & 37                & 11           & 12              & 82.35\%                  \\ \hline
Teleoperation                                & 7                 & 21                & 1            & 1               & 96.67\%                  \\ \hline
Speech Generation                            & 2                 & 1                 & 5            & 1               & 88.89\%                  \\ \hline
Switch                                       & 3                 & 2                 & 1            & 0               & 100.00\%                 \\ \hline
Mobile                                       & 2                 & 26                & 4            & 6               & 84.21\%                  \\ \hline
Manipulator                                  & 2                 & 30                & 2            & 3               & 91.89\%                  \\ \hline
Speaker                                      & 0                 & 0                 & 5            & 1               & 83.33\%                  \\ \hline
Sensor                                       & 10                & 77                & 16           & 25              & 80.47\%                  \\ \hline
Visualization                                & 98                & 60                & 11           & 0               & 100.00\%                 \\ \hline
Support                                      & 14                & 45                & 25           & 27              & 75.68\%                  \\ \hline
Extension                                    & 23                & 72                & 18           & 106             & 51.60\%                  \\ \hline
\textcolor{red}{\textbf{Function Node}}      & 154               & 335               & 74           & 73              & 88.52\%                  \\ \hline
\end{tabular}}
\label{table:automation}
\vspace{-15pt}
\end{table}

\section{The Description of ROS Platform and Apps}
\label{sec:robot-platform-app-describe}
\subsection{The Relationship of ROS Elements}
\label{sec:repo-func-pkg}
\begin{figure}[tb]
\centering
\includegraphics[width=1.00\linewidth]{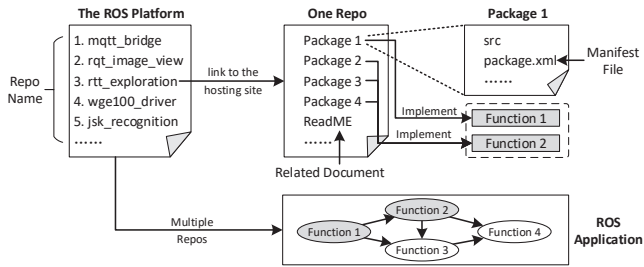}
\caption{The relationship among the app, repo, package and function.}
\label{fig:repo-func-pkg}
\end{figure}

Figure \ref{fig:repo-func-pkg} shows the relationship of different elements in the ROS platform. 
As described in \cref{sec:2-ros}, the ROS platforms maintains a list of ROS indexes (i.e. repo names), each index link to the source code of this repo in the hosting site. 
A repo commonly consists of one or multiple ROS packages, and its related document (i.e. ReadMe). Each package concludes source code and a manifest file (i.e. package.xml). 
The manifest file is used to describe its version, description and dependence. 
A robotic function can be implemented by one or multiple packages. It means one repo can have two or more functions. 
These functions are then integrated with some functions in other repos or customized by users to consist a ROS application. 

\subsection{Descriptions of Commonly Used Apps}
\label{sec:robot-app-describe}
We describe the top five robot apps in ROS platform that are most commonly used, according to Table \ref{table:apps}.

\noindent
\textbf{Remote control.} 
This type of apps is designed to control the robot remotely from smartphones, joysticks or keyboards. It uses \texttt{Teleoperration} to receive the signals from the remote controller and \texttt{Mobile/Manipulator Driver} to transfer these signals to each actuator's control command.

\noindent
\textbf{2D/3D mapping.} 
These apps aim to create a 2D/3D map of unknown environment through remote control. End users use \texttt{Tele\\operration} to move the robot to explore the unknown zones. During the exploration, \texttt{Preprocessing} sends structural sensory data to \texttt{Mapping} for map creation.

\noindent
\textbf{Navigation.} 
This type of apps instructs a robot to navigate through an obstacle-filled known environment and reach a specified destination. These apps use \texttt{Localization} to estimate the robot's position,  and \texttt{Path Planning} to compute a collision-free path from its position to the destination. Then, \texttt{Path Tracking} is called to follow the path until the robot achieves the goal or the mission  fails. 

\noindent
\textbf{SLAM.} 
These apps can be regarded as the combination of Mapping and Navigation. To reach an arbitrary destination in an obstacle-filled unknown environment, the apps use \texttt{Mapping} and \texttt{Localization} simultaneously to transfer the unknown map to a known one and locate its position.

\noindent
\textbf{Face/Person Detection.} 
These apps receive images from cameras (\texttt{Preprocessing}) and apply the OpenCV face/person detector based on an Adaboost cascade of Haar features/HOG (\texttt{Recognition}). They publish regions of interests (ROIs) of the detection and a debug image, showing the processed image with the ROIs that is likely to contain faces or persons.

\section{Mission-specific Risk Message Type}
\label{sec:A-msg-type}
We present the description of the EVENT\_MSG\_TYPE in Table \ref{table:event_msg_type} and ACTION\_MSG\_TYPE in Table \ref{table:action_msg_type}.

\makeatletter
\setlength{\@fptop}{0pt}
\setlength{\@fpbot}{0pt plus 1fil}
\makeatother

\begin{table}[tb]
\caption{Description of EVENT\_MSG\_TYPE.}
\centering
\resizebox{1\linewidth}{!}{
\begin{tabular}{|l|l|}
\hline
\rowcolor{gray!40} 
\textbf{Message Type}   & \multicolumn{1}{l|}{\textbf{Description}}  \\
\hline
\makecell[l]{sensor\_msgs/\\BatteryState} & \makecell[l]{Measurement of the battery state (voltage, \\charge, etc).}  \\
\hline
\makecell[l]{sensor\_msgs/\\Temperature} & Measurement of the temperature.  \\
\hline
\makecell[l]{sensor\_msgs/\\RelativeHumidity} & \makecell[l]{Defines the ratio of partial pressure of water \\vapor to the saturated vapor pressure at a \\temperature.} \\
\hline
\makecell[l]{sensor\_msgs/\\MagneticField} & \makecell[l]{Measurement of the Magnetic Field vector at \\a specific location.}  \\
\hline
\makecell[l]{sensor\_msgs/\\FluidPressure} & \makecell[l]{Measurement of the pressure inside of a fluid \\(air, water, etc), atmospheric or barometric \\pressure.}  \\
\hline
\makecell[l]{sensor\_msgs/\\NavSatFix} & \makecell[l]{Measurement for any Global Navigation \\Satellite System (latitude, longitude, etc).}  \\
\hline
\makecell[l]{sensor\_msgs/\\Illuminance} & \makecell[l]{Measurement of the single photometric \\illuminance.}  \\
\hline
\makecell[l]{nav\_msgs/\\Odometry} & \makecell[l]{Measurement of an estimate of a position \\and velocity in free space (pose, twist, etc).}  \\
\hline
\end{tabular}}
\label{table:event_msg_type}
\end{table}

\begin{table}[tb]
\caption{Description of ACTION\_MSG\_TYPE.}
\centering
\resizebox{1\linewidth}{!}{
\begin{tabular}{|c|l|l|}
\hline
\rowcolor{gray!40} 
\textbf{Actuator} & \textbf{Message Type} & \multicolumn{1}{c|}{\textbf{Description}} \\
\hline
Mobile & \makecell[l]{geometry\_msg/\\Twist} & \makecell[l]{This expresses the velocity in \\free space broken into its linear \\and angular parts.}  \\
\hline
Manipulator	&\makecell[l]{control\_msgs/\\FollowJoint\\TrajectoryAction} & \makecell[l]{This defines the joint trajectory \\to follow.}  \\
\hline
Speaker	& \makecell[l]{audio\_common\\\_msg/AudioData} & \makecell[l]{This defines the audio data to \\speak.}  \\
\hline
\end{tabular}}
\label{table:action_msg_type}
\end{table}

\section{The \textsc{RTron} User Console}
\label{sec:B-user-console}
\begin{figure*}[tb]
\centering
\includegraphics[width=1.00\linewidth]{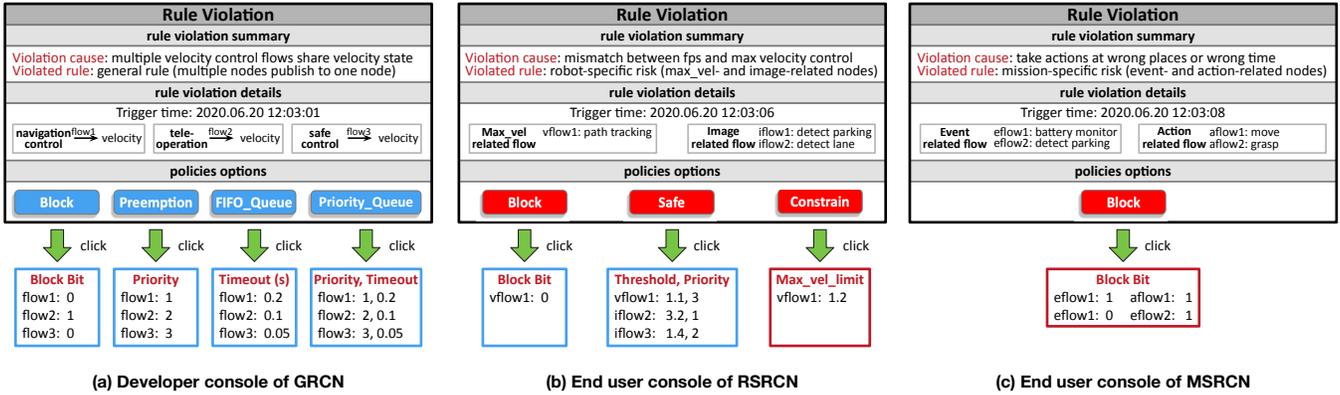}
\caption{Developer and end user console of each risk in \textsc{RTron}. The red solid rectangle denotes a button for the end users. The blue/red box represents policy-related configuration parameters for the developers/end users.}
\label{fig:6-console}
\end{figure*}

We present the description of the end user console of general risk coordination node, robot-specific risk coordination node and mission-specific risk coordination node in Figure \ref{fig:6-console}(a), Figure \ref{fig:6-console}(b) and Figure \ref{fig:6-console}(c), respectively.

\section{The App Interaction Graph}
\label{sec:C-app-graph}

\begin{table*}[tb]
\caption{Examples of high-risk nodes in the Home and AutoRace apps. Texts marked in red are for risk identification in our system.}
\centering
\resizebox{1\linewidth}{!}{
\begin{tabular}{|c|l|l|l|l|l|l|l|} 
\hline
\rowcolor{gray!40} 
\textbf{Scenario} & 
\multicolumn{1}{c|}{\textbf{Risk Type}} & \multicolumn{1}{c|}{\textbf{High-Risk Nodes}} & \multicolumn{1}{c|}{\textbf{Sub Topic Name}} & \multicolumn{1}{c|}{\textbf{Sub Topic Type}} & \multicolumn{1}{c|}{\textbf{Pub Topic Name}} & \multicolumn{1}{c|}{\textbf{Pub Topic Type}} & \multicolumn{1}{c|}{\textbf{Pub Node}}  \\ 
\hline
 \multirow{7}{*}{Home} & \multirow{2}{*}{GR-ST} & /move\_base & - & - & \textcolor{red}{/cmd\_vel} & geometry\_msgs/Twist & /gazebo \\ 
 \cline{3-8}
  & & /teleop\_twist\_keyboard & - & - & \textcolor{red}{/cmd\_vel} & geometry\_msgs/Twist & /gazebo \\ 
 \cline{2-8}
  & \multirow{2}{*}{GR-MT} & /gazebo & - & - & /camera/depth/image\_raw & \textcolor{red}{sensor\_msgs/Image} & \textcolor{red}{/find\_object\_3d} \\ 
 \cline{3-8}
  & & /gazebo & - & - & /camera/rgb/image\_raw & \textcolor{red}{sensor\_msgs/Image} & \textcolor{red}{/find\_object\_3d} \\ 
 \cline{2-8}
  & RSR-Image & /find object 3d & /camera/rgb/image raw & \textcolor{red}{sensor\_msgs/Image} & \textcolor[rgb]{0.984,0,0.027}{/objects} & std msgs/Float32MultiArray & /search manager \\ 
 \cline{2-8}
  & MSR-Event & /move\_base & /odom & \textcolor{red}{nav\_msgs/Odometry} & - & - & - \\ 
 \cline{2-8}
  & MSR-Action & /rosbot\_tts & - & - & audio\_common\_msgs/AudioData & \textcolor{red}{/rosbot\_audio/audio} & /rosbot\_audio \\ 
 \hline
\multirow{8}{*}{AutoRace} & \multirow{2}{*}{GR-ST} & /detect\_tunnel & - & - & \textcolor{red}{/move\_base\_simple/goal} & geometry\_msgs/PoseStamped & /move\_base\_simple/goal \\ 
\cline{3-8}
& & /rviz & - & - & \textcolor{red}{/move\_base\_simple/goal} & geometry\_msgs/PoseStamped & /move\_base\_simple/goal \\ 
\cline{2-8}
& \multirow{2}{*}{GR-MT} & /detect/lane & - & - & /detect/lane & \textcolor{red}{std\_msgs/Float64} & \textcolor{red}{/control/lane} \\ 
\cline{3-8}
& & /detect\_traffic\_light & - & - & /control/max\_vel & \textcolor{red}{std\_msgs/Float64} & \textcolor{red}{/control/lane} \\ 
\cline{2-8}
& RSR-Image & /detect\_sign & /camera/image\_compensated & \textcolor{red}{sensor\_msgs/Image} & \textcolor{red}{/detect/traffic\_sign} & std\_msgs/UInt8 & /core\_mode\_decider \\ 
\cline{2-8}
& RSR-Max\_vel & /detect\_parking & - & - & \textcolor{red}{/control/max\_vel} & \textcolor{red}{std\_msgs/Float64} & /control\_lane \\ 
\cline{2-8}
& MSR-Event & /core\_node\_controller & \textcolor{red}{/detect/tunnel\_stamped} & std\_msgs/UInt8 & - & - & - \\ 
\cline{2-8}
& MSR-Action & /detect\_tunnel & - & - & /cmd\_vel & \textcolor{red}{geometry\_msgs/Twist} & /gazebo \\
\hline                             
\end{tabular}}
\label{table:risk_examples}
\vspace{-10pt}
\end{table*}

\begin{figure*}[tb]
\centering
\includegraphics[width=0.7\linewidth]{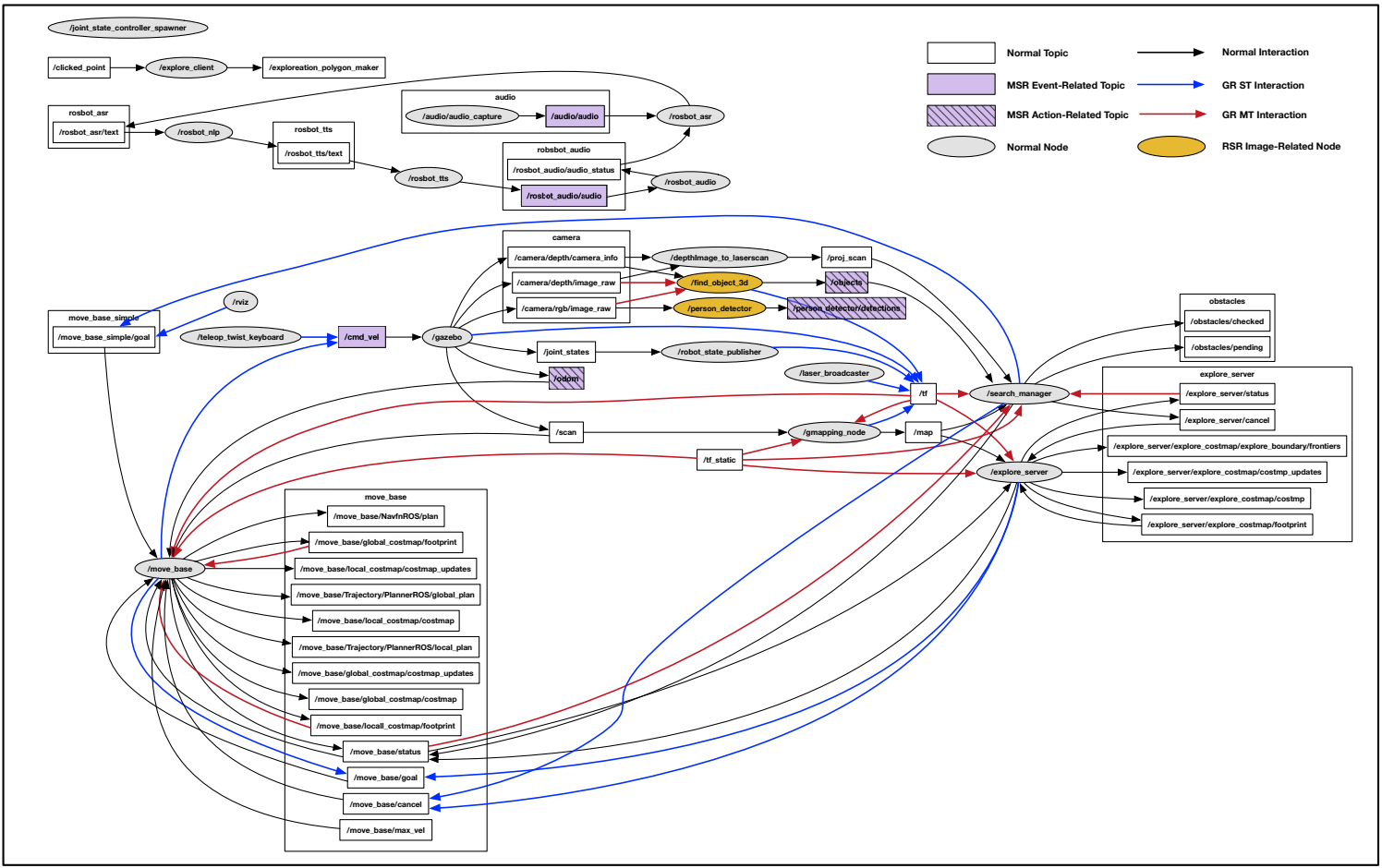}
\caption{The Interaction Graph of Robot Applications in Home Scenario. The subscriptions of visualization node (i.e. /rviz) and log node (i.e. /rosnode) are deleted in the figure.}
\label{fig:C-home-rosgraph}
\end{figure*}

Figure \ref{fig:C-home-rosgraph} shows a complete interaction graph of home app. Gray ellipses denote the function nodes; while rectangles represent topics. Each two nodes are connected through topics. We use black arrows to denote these interactions. The interactions of GR-ST and GR-MT are marked with blue and red, receptively. We also use purple rectangles with/without diagonal stripes to denote MSR event-related and MSR action-related topics. The RSR image-related nodes is depicted in yellow ellipses. Table \ref{table:risk_examples} gives one example of the identified high-risk node for each type in two apps. 

\begin{figure*}[tb]
\centering
\includegraphics[width=0.7\linewidth]{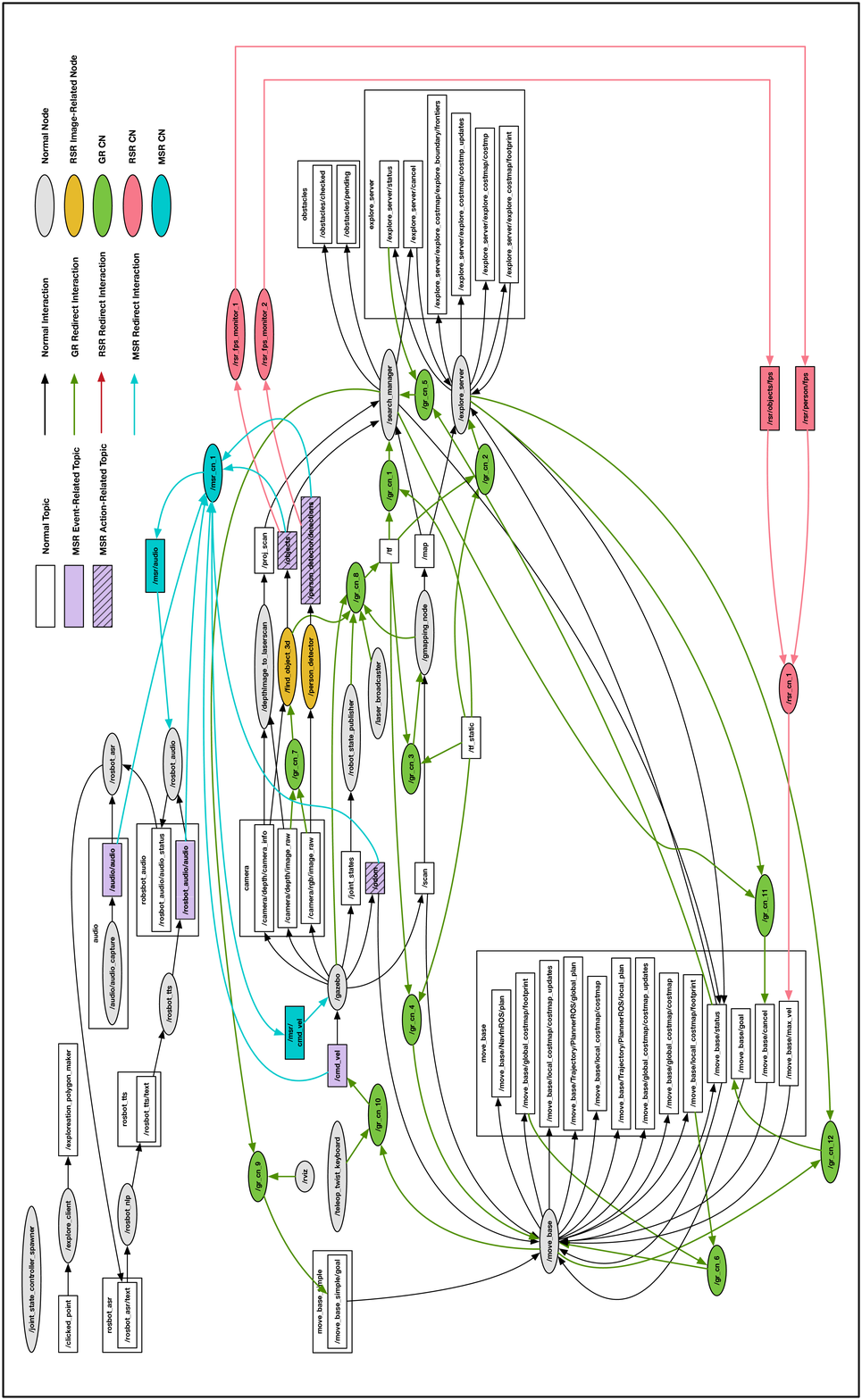}
\caption{The Coordination Node Distribution in the Interaction Graph of Robot Applications in Home Scenario.}
\label{fig:C-home-cn-rosgraph}
\end{figure*}

Figure \ref{fig:C-home-cn-rosgraph} shows the coordination node distribution in the complete interaction graph of home app. We use green, red and cyan ellipses to denote GR, RSR and MSR CN receptively. The redirected interactions between normal/high-risk nodes and CN are marked with the same color of each type of CN.

\begin{figure}[tb]
\centering
\includegraphics[width=1.0\linewidth]{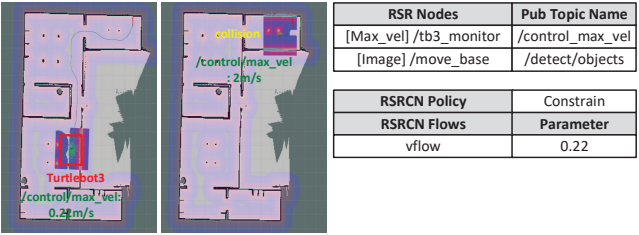}
\caption{RSR experiment on Turtlebot3 in the simulation.}
\label{fig:C-rsr-case}
\vspace{-10pt}
\end{figure}

\lstset{style=mystyle-2}
\begin{figure*}[h]
  \centering
  \removelatexerror
  \begin{minipage}[b]{.31\linewidth}
\begin{lstlisting}[language=Python]
def gr_attack():
  if mac == HOST_MAC:
    if cur_time_min == 15 and 
       node_exist('move_base'):

      twist = Twist()
      twist.linear.x = twist.linear.y      = twist.linear.z = 0.0
      twist.angular.x = twist.angular      .y = 0.0
      twist.angular.z = -0.2

      gr_atk_pub = rospy.Publish(         'cmd_vel', Twist,                queue_size=10)
      gr_atk_pub.publish(twist)
\end{lstlisting}
  \subcaption{\footnotesize{Malicious codes of GR attack}}
  \label{fig:gr-atk}
  \end{minipage} \hfill
  \begin{minipage}[b]{.31\linewidth}
\begin{lstlisting}[language=Python]
def rsr_attack():
  if mac == HOST_MAC:
    if cur_time_min == 30 and 
       topic_exist('control/max_vel'):

      max_vel = Float64()
      max_vel.data = 2

      rsr_atk_pub = rospy.Publisher(
      		'control/max_vel', Float64, 
      		queue_size=1)
      rsr_atk_pub.publish(max_vel)
\end{lstlisting}
  \subcaption{\footnotesize{Malicious codes of RSR attack}}
  \label{fig:rsr-atk}
  \end{minipage} \hfill
  \begin{minipage}[b]{.31\linewidth}
\begin{lstlisting}[language=Python]
def msr_attack():
  if mac == HOST_MAC:
    if cur_time_min == 45 and 
       topic_exist('move_base_simple/goal'):

      goal = PoseStamped()
      goal.header.stamp = now()
      goal.header.frame_id = "map"
      goal.pose.position = MAL_LOC

      msr_atk_pub = rospy.Publisher(
      		'move_base_simple/goal', 
      		PoseStamped, queue_size=1)
      msr_atk_pub.publish(goal)
\end{lstlisting}
  \subcaption{\footnotesize{Malicious codes of MSR attack}}
  \label{fig:msr-atk}
  \end{minipage} \hfill
  \caption{Malicious codes in our \texttt{tb3\_safe\_teleop} and \texttt{tb3\_monitor} packages.}
  \label{fig:malicious-codes}
\end{figure*}

\section{The Attack Implementation and RSR Case}
\label{sec:D-exp}
\subsection{The Attack Implementation}
\label{sec:D-attack-impl}
As \cref{sec:7-4-usecase} describes, the malicious codes of three types of risks are hidden in the two normal robot functional packages: \texttt{/tb3\_safe\_control} and \texttt{tb3\_monitor}. The \texttt{/tb3\_safe\_control} provides commands for safe teleoperation with different input devices. It use LaserScan information to estimated the distance between the robot and obstacles, and stop the robot's movement within a customized safe distance. The \texttt{tb3\_monitor} package provides commands to monitor nodes' information and robot's states in real time.

Figure \ref{fig:malicious-codes} illustrates the attack and its consequences. The malicious code of GR attack is added in the \texttt{/tb3\_safe\_control} package. The malicious codes of RSR and MSR attack are all hidden in the \texttt{tb3\_monitor} package. It's worth noting that we add specific triggering logics (Lines 2) in each attack to avoid raising ethical concerns. The triggering condition is the success match between the default MAC address and local host MAC address. Since the MAC address is unique of different devices, the malicious codes can only work in our robotic devices. Moreover, we set time matching process to make the attack launch at specific time, other than at the beginning. This can make attacks more hidden.

Figure \ref{fig:malicious-codes}(a) shows the related code snippets in the GR attack.The gr\_attack function is invoked by a callback function of LaserScan Topic. In each iteration, the function starts by searching the gr-related vulnerable node, i.e. `\texttt{move\_base}'. If exists, it means the move-related control topic `\texttt{cmd\_vel}' exists and the robot is executing a navigation task with a great probability. Thus, we send a Twist-type move command with -0.2 z-axis angular velocity to the victim topic. This would lead to the robot suddenly turn right and crash to the obstacles while navigating in a collision-free path.

Figure \ref{fig:malicious-codes}(b) and (c) present the malicious code snippets in the RSR and MSR attack respectively. Both rsr\_attack and msr\_attack functions are invoked while each traversal of all topics of one node. Specifically, once the `\texttt{control/max\_vel}' topic exists, the malicious process would send a max velocity control command with 2 m/s to the victim topic. Similarly, if the `\texttt{move\_base\_simple/goal}' topic exists, a goal with a malicious location will be launched to the victim topic and the robot would move to the dangerous destination.

\subsection{The RSR Case Study in Simulation}
\label{sec:D-rsr-exp}
We use the same simulation setup described in Section \ref{sec:7-evaluation}. 
The \texttt{tb3\_monitor} node sends malicious max velocity configuring commands druing the robot's navigation at specific moments. It increases the max velocity value through publishing the malicious messages to the \texttt{/control\_max\_vel} topic. In Figure \ref{fig:C-rsr-case}, the initial max velocity is 0.22m/s and the robot moves safely. At one moment, this value is increased to 2m/s. Then the robot moves too fast to detect the obstacle and a collision occurs. In \textsc{RTron}, we choose \emph{Constrain} policy in the RSRCN and set 0.22 to the vflow. In this way, the max velocity of the robot is fixed at 0.22m/s and cannot be changed by the attacker.

%% file: master.bbl
\begin{thebibliography}{10}
\providecommand{\url}[1]{#1}
\csname url@samestyle\endcsname
\providecommand{\newblock}{\relax}
\providecommand{\bibinfo}[2]{#2}
\providecommand{\BIBentrySTDinterwordspacing}{\spaceskip=0pt\relax}
\providecommand{\BIBentryALTinterwordstretchfactor}{4}
\providecommand{\BIBentryALTinterwordspacing}{\spaceskip=\fontdimen2\font plus
\BIBentryALTinterwordstretchfactor\fontdimen3\font minus
  \fontdimen4\font\relax}
\providecommand{\BIBforeignlanguage}[2]{{%
\expandafter\ifx\csname l@#1\endcsname\relax
\typeout{** WARNING: IEEEtran.bst: No hyphenation pattern has been}%
\typeout{** loaded for the language `#1'. Using the pattern for}%
\typeout{** the default language instead.}%
\else
\language=\csname l@#1\endcsname
\fi
#2}}
\providecommand{\BIBdecl}{\relax}
\BIBdecl

\bibitem{openxc}
``Openxc platform,'' \url{http://openxcplatform.com/}, 2020.

\bibitem{onboard-sdk}
``Dji onboard sdk,'' \url{https://developer.dji.com/onboard-sdk/}, 2020.

\bibitem{ur-app-builder}
``Application builder,'' \url{https://www.universal-robots.com/builder/}, 2020.

\bibitem{ros}
``Open source robot operating system,'' \url{http://www.ros.org/}, 2019.

\bibitem{ros-pr2}
``Ros pr2 package,'' \url{http://wiki.ros.org/Robots/PR2/}, 2020.

\bibitem{ros-abb}
``Ros abb package,'' \url{http://wiki.ros.org/abb/}, 2020.

\bibitem{android-store}
``Google play,'' \url{https://play.google.com/store/}, 2020.

\bibitem{ios-store}
``App store,'' \url{https://www.apple.com/ios/app-store/}, 2020.

\bibitem{windows-store}
``Windows apps - microsoft store,''
  \url{https://www.microsoft.com/en-us/store/apps/windows/}, 2020.

\bibitem{ubuntu-store}
``Ubuntu appstore,'' \url{https://ubuntu.com/blog/tag/appstore/}, 2020.

\bibitem{mac-store}
``The mac app store,'' \url{https://www.apple.com/uk/osx/apps/app-store//},
  2020.

\bibitem{smarthings}
``Samsung smartthings,'' \url{https://www.smartthings.com/}, 2020.

\bibitem{homekit}
``Apple homekit,'' \url{https://developer.apple.com/homekit/}, 2020.

\bibitem{weave}
``Google weave project,'' \url{https://developers.google.com/weave/}, 2020.

\bibitem{ros-scan}
N.~DeMarinis, S.~Tellex, V.~P. Kemerlis, G.~D. Konidaris, and R.~Fonseca,
  ``Scanning the internet for ros: A view of security in robotics research,''
  in \emph{International Conference on Robotics and Automation (ICRA)}, 2019.

\bibitem{sec-analysis2}
D.~Quarta, M.~Pogliani, M.~Polino, A.~M. Zanchettin, and S.~Zanero, ``Rogue
  robots: Testing the limits of an industrial robot’s security,'' Politecnico
  di Milano, Tech. Rep., 2017.

\bibitem{chrysler}
``After jeep hack, chrysler recalls 1.4m vehicles for bugfix,''
  \url{https://www.wired.com/2015/07/jeep-hack-chrysler-recalls-1-4m-vehicles-bug-fix/},
  2015.

\bibitem{iot-sift}
C.-J.~M. Liang, B.~F. Karlsson, N.~D. Lane, F.~Zhao, J.~Zhang, Z.~Pan, Z.~Li,
  and Y.~Yu, ``Sift: building an internet of safe things,'' in
  \emph{International Symposium on Information Processing in Sensor Networks
  (IPSN)}, 2015.

\bibitem{iot-salus}
C.-J.~M. Liang, Z.~L. Lei~Bu, J.~Zhang, S.~Han, B.~F. Karlsson, D.~Zhang, and
  F.~Zhao, ``Systematically debugging iot control system correctness for
  building automation,'' in \emph{Proceedings of the 3rd ACM International
  Conference on Systems for Energy-Efficient Built Environments
  (BuildSys@SenSys)}, 2015.

\bibitem{iot-iotMon}
W.~Ding and H.~Hu, ``On the safety of iot device physical interaction
  control,'' in \emph{ACM Conference on Computer and Communications Security
  (CCS)}, 2018.

\bibitem{iot-iotGuard}
Z.~B. Celik, G.~Tan, and P.~D. McDaniel, ``Iotguard: Dynamic enforcement of
  security and safety policy in commodity iot,'' in \emph{Annual Network and
  Distributed System Security Symposium (NDSS)}, 2019.

\bibitem{iot-soteria}
Z.~B. Celik, P.~D. McDaniel, and G.~Tan, ``Soteria: Automated iot safety and
  security analysis,'' in \emph{Annual Technical Conference (ATC)}, 2018.

\bibitem{iot-iostSan}
D.~T. Nguyen, C.~Song, Z.~Qian, S.~V. Krishnamurthy, E.~J.~M. Colbert, and
  P.~D. McDaniel, ``Iotsan: fortifying the safety of iot systems,'' in
  \emph{Conference on Emerging Network Experiment and Technology (CoNEXT)},
  2018.

\bibitem{iot-homeGuard}
H.~Chi, Q.~Zeng, X.~Du, and J.~Yu, ``Cross-app interference threats in smart
  homes: Categorization, detection and handling,'' in \emph{CoRR
  abs/1808.02125}, 2018.

\bibitem{iot-menShen}
L.~Bu, W.~Xiong, C.-J.~M. Liang, S.~Han, D.~Zhang, S.~Lin, and X.~Li,
  ``Systematically ensuring the confidence of real-time home automation iot
  systems,'' \emph{ACM Transactions on Cyber-Physical Systems (TCPS)}, vol.~2,
  no.~3, pp. 22:1--22:23, 2018.

\bibitem{iot-autoTap}
L.~Zhang, W.~He, J.~Martinez, N.~Brackenbury, S.~Lu, and B.~Ur, ``Autotap:
  synthesizing and repairing trigger-action programs using ltl properties,'' in
  \emph{International Conference on Software Engineering (ICSE)}, 2019.

\bibitem{iot-iRuler}
Q.~Wang, P.~Datta, W.~Yang, S.~Liu, A.~Bates, and C.~A. Gunter, ``Charting the
  attack surface of trigger-action iot platforms,'' in \emph{ACM Conference on
  Computer and Communications Security (CCS)}, 2019.

\bibitem{app-autoware}
``The autoware.ai project,'' \url{https://github.com/Autoware-AI/autoware.ai},
  2020.

\bibitem{app-apollo}
``Baidu apollo,'' \url{https://github.com/ApolloAuto/apollo}, 2020.

\bibitem{handbook}
Siciliano, Bruno, and O.~Khatib, \emph{Springer handbook of robotics}.\hskip
  1em plus 0.5em minus 0.4em\relax Secaucus, NJ, USA: Sprinter-Verlag New York,
  Inc.: Springer, 2016.

\bibitem{pr}
S.~Thrun, W.~Burgard, and D.~Fox, \emph{Probabilistic Robotics}.\hskip 1em plus
  0.5em minus 0.4em\relax The MIT Press, 2005.

\bibitem{ros-vulnerable}
B.~Dieber, B.~Breiling, S.~Taurer, S.~Kacianka, S.~Rass, and P.~Schartner,
  ``Security for the robot operating system,'' \emph{IEEE Trans. Robotics and
  Autonomous Systems}, vol.~98, pp. 192--203, 2017.

\bibitem{robot-vulnerable}
T.~Denning, C.~Matuszek, K.~Koscher, J.~R. Smith, and T.~Kohno, ``A spotlight
  on security and privacy risks with future household robots: attacks and
  lessons,'' in \emph{Ubiquitous Computing (UbiComp)}, 2009.

\bibitem{hackingrobots}
C.~Cerrudo and L.~Apa, ``Hacking robots before skynet,'' \emph{IOActive
  Website}, 2017.

\bibitem{ros-3rd-party-threat}
``Ros 2 robotic systems threat model,''
  \url{https://design.ros2.org/articles/ros2_threat_model.html}, 2020.

\bibitem{ros-assessment}
J.~R. Mcclean and C.~Farrar, ``A preliminary cyber-physical security assessment
  of the robot operating system (ros),'' in \emph{Proceedings of SPIE}, 2013.

\bibitem{rvd}
``Robot vulnerability database (rvd),''
  \url{https://github.com/aliasrobotics/RVD/}, 2020.

\bibitem{ros-authen-1}
R.~Toris, C.~A. Shue, and S.~Chernova, ``Message authentication codes for
  secure remote non-native client connections to ros enabled robots,'' in
  \emph{International Conference on Technologies for Practical Robot
  Applications (TePRA)}, 2014.

\bibitem{ros-authen-2}
R.~Dóczi, B.~S. Ferenc~Kis, V.~Poser, G.~Kronreif, E.~Josvai, and
  M.~Kozlovszky, ``Increasing ros 1.x communication security for medical
  surgery robot,'' in \emph{IEEE International Conference on Systems, Man and
  Cybernetics (SMC)}, 2016.

\bibitem{ros-authen-3}
B.~Dieber, S.~Kacianka, S.~Rass, and P.~Schartner, ``Application-level security
  for ros-based applications,'' in \emph{International Conference on
  Intelligent RObots and Systems (IROS)}, 2016.

\bibitem{crypto-1}
V.~Matellán, J.~Balsa, F.~Casado, C.~Fernández, and F.~J.~R. Lera,
  ``Cybersecurity in autonomous systems: Evaluating the performance of
  hardening ros,'' in \emph{XVII Workshop En Agentes Físicos}, 2016.

\bibitem{crypto-2}
T.~Bonaci, J.~Herron, T.~Yusuf, J.~Yan, T.~Kohno, and H.~J. Chizeck, ``To make
  a robot secure: An experimental analysis of cyber security threats against
  teleoperated surgical robots,'' in \emph{CoRR abs/1504.04339}, 2015.

\bibitem{crypto-3}
N.~M. Rodday, R.~de~Oliveira~Schmidt, and A.~Pras, ``Exploring security
  vulnerabilities of unmanned aerial vehicles,'' in \emph{IEEE/IFIP Network
  Operations and Management Symposium (NOMS)}, 2016.

\bibitem{dos-2}
J.~Chen, Z.~Feng, J.-Y. Wen, B.~Liu, and L.~Sha, ``A container-based dos
  attack-resilient control framework for real-time uav systems,'' in
  \emph{Design, Automation, and Test in Europe (DATE)}, 2019.

\bibitem{spoof-gps1}
N.~O. Tippenhauer, C.~Pöpper, K.~B. Rasmussen, and S.~Capkun, ``On the
  requirements for successful gps spoofing attacks,'' in \emph{ACM Conference
  on Computer and Communications Security (CCS)}, 2011.

\bibitem{spoof-gps2}
N.~Nighswander, B.~M. Ledvina, J.~Diamond, R.~Brumley, and D.~Brumley, ``Gps
  software attacks,'' in \emph{ACM Conference on Computer and Communications
  Security (CCS)}, 2012.

\bibitem{spoof-gps3}
K.~C. Zeng, S.~Liu, Y.~Shu, D.~Wang, H.~Li, Y.~Dou, G.~Wang, and Y.~Yang, ``All
  your gps are belong to us: Towards stealthy manipulation of road navigation
  systems,'' in \emph{USENIX Security Symposium (USENIX Security 18)}, 2018.

\bibitem{spoof-gps4}
J.~S. Warner and R.~G. Johnston, ``A simple demonstration that the global
  positioning system (gps) is vulnerable to spoofing,'' \emph{Journal of
  Security Administration}, vol.~25, no.~2, pp. 19--27, 2002.

\bibitem{spoof-gps5}
S.-H. Seo, B.-H. Lee, S.-H. Im, and G.-I. Jee, ``Effect of spoofing on unmanned
  aerial vehicle using counterfeited gps signal,'' \emph{Journal of
  Positioning, Navigation, and Timing}, vol.~4, no.~2, p. 57–65, 2015.

\bibitem{spoof-lidar-1}
H.~Shin, D.~Kim, Y.~Kwon, and Y.~Kim, ``Illusion and dazzle: Adversarial
  optical channel exploits against lidars for automotive applications,'' in
  \emph{International Workshop on Cryptographic Hardware and Embedded Systems
  (CHES)}, 2017.

\bibitem{spoof-lidar-2}
Y.~Cao, C.~Xiao, B.~Cyr, Y.~Zhou, W.~Park, S.~Rampazzi, Q.~A. Chen, K.~Fu, and
  Z.~M. Mao, ``Adversarial sensor attack on lidar-based perception in
  autonomous driving,'' in \emph{ACM Conference on Computer and Communications
  Security (CCS)}, 2019.

\bibitem{spoof-optical}
D.~Davidson, H.~Wu, R.~Jellinek, V.~Singh, and T.~Ristenpart, ``Controlling
  uavs with sensor input spoofing attacks,'' in \emph{Workshop on Offensive
  Technologies (WOOT)}, 2016.

\bibitem{spoof-gyroscopic-1}
Y.~Son, H.~Shin, D.~Kim, Y.-S. Park, J.~Noh, K.~Choi, J.~Choi, and Y.~Kim,
  ``Rocking drones with intentional sound noise on gyroscopic sensors,'' in
  \emph{USENIX Security Symposium (USENIX Security 15)}, 2015.

\bibitem{spoof-gyroscopic-2}
T.~Trippel, O.~Weisse, W.~Xu, P.~Honeyman, and K.~Fu, ``Walnut: Waging doubt on
  the integrity of mems accelerometers with acoustic injection attacks,'' in
  \emph{EuroS\&P}, 2017.

\bibitem{spoof-gyroscopic-3}
Y.~Tu, Z.~Lin, I.~Lee, and X.~Hei, ``Injected and delivered: Fabricating
  implicit control over actuation systems by spoofing inertial sensors,'' in
  \emph{USENIX Security Symposium (USENIX Security 18)}, 2018.

\bibitem{inject-magnetic}
Y.~Shoukry, P.~D. Martin, P.~Tabuada, and M.~B. Srivastava, ``Non-invasive
  spoofing attacks for anti-lock braking systems,'' in \emph{International
  Workshop on Cryptographic Hardware and Embedded Systems (CHES)}, 2013.

\bibitem{spoof-actuator}
F.~Fei, Z.~Tu, R.~Yu, T.~Kim, X.~Zhang, D.~Xu, and X.~Deng, ``Cross-layer
  retrofitting of uavs against cyber-physical attacks,'' in \emph{IEEE
  International Conference on Robotics and Automation (ICRA)}, 2018.

\bibitem{spoof-signal}
H.~Choi, W.-C. Lee, Y.~Aafer, F.~Fei, Z.~Tu, X.~Zhang, D.~Xu, and X.~Xinyan,
  ``Detecting attacks against robotic vehicles: A control invariant approach,''
  in \emph{ACM Conference on Computer and Communications Security (CCS)}, 2018.

\bibitem{sec-analysis}
D.~Quarta, M.~Pogliani, M.~Polino, F.~Maggi, A.~M. Zanchettin, and S.~Zanero,
  ``An experimental security analysis of an industrial robot controller,'' in
  \emph{IEEE Symposium on Security and Privacy (S\&P)}, 2017.

\bibitem{nlp-1}
X.~Pan, Y.~Cao, X.~Du, B.~He, G.~Fang, R.~Shao, and Y.~Chen, ``Flowcog:
  Context-aware semantics extraction and analysis of information flow leaks in
  android apps,'' in \emph{USENIX Security Symposium (USENIX Security 18)},
  2018.

\bibitem{nlp-2}
E.~Fernandes, J.~Jung, and A.~Prakash, ``Security analysis of emerging smart
  home applications,'' in \emph{IEEE Symposium on Security and Privacy (S\&P)},
  2016.

\bibitem{nlp-3}
R.~Xu, H.~Saïdi, and R.~J. Anderson, ``Aurasium: Practical policy enforcement
  for android applications,'' in \emph{USENIX Security Symposium (USENIX
  Security 12)}, 2012.

\bibitem{nlp-4}
X.~yong Zhou, S.~Demetriou, D.~He, M.~Naveed, X.~Pan, X.~Wang, C.~A. Gunter,
  and K.~Nahrstedt, ``Identity, location, disease and more: inferring your
  secrets from android public resources,'' in \emph{ACM Conference on Computer
  and Communications Security (CCS)}, 2013.

\bibitem{nlp-5}
E.~Fernandes, J.~Paupore, A.~Rahmati, D.~Simionato, M.~Conti, and A.~Prakash,
  ``Flowfence: Practical data protection for emerging iot application
  frameworks,'' in \emph{USENIX Security Symposium (USENIX Security 16)}, 2016.

\bibitem{nlp-6}
Y.~M.~P. Pa, S.~Suzuki, K.~Yoshioka, T.~Matsumoto, T.~Kasama, and C.~Rossow,
  ``Iotpot: Analysing the rise of iot compromises,'' in \emph{USENIX Workshop
  on Offensive Technologies (WOOT)}, 2015.

\bibitem{nlp-7}
X.~yong Zhou, Y.~Lee, N.~Zhang, M.~Naveed, and X.~Wang, ``The peril of
  fragmentation: Security hazards in android device driver customizations,'' in
  \emph{IEEE Symposium on Security and Privacy (S\&P)}, 2014.

\bibitem{tokens-regex}
``Stanford tokensregex,''
  \url{https://nlp.stanford.edu/software/tokensregex.html}, 2020.

\bibitem{ros-robot}
``Robots that you can use with ros.'' \url{https://robots.ros.org/}, 2020.

\bibitem{ros-sensors}
``Sensors supported by ros.'' \url{http://wiki.ros.org/Sensors/}, 2020.

\bibitem{mavbench}
B.~Boroujerdian, H.~Genc, S.~Krishnan, W.~Cui, A.~Faust, and V.~J. Reddi,
  ``Mavbench: Micro aerial vehicle benchmarking,'' in \emph{Annual IEEE/ACM
  International Symposium on Microarchitecture (MICRO)}, 2018.

\bibitem{autovehicle}
S.-C. Lin, Y.~Zhang, C.-H. Hsu, M.~Skach, M.~E. Haque, L.~Tang, and J.~Mars,
  ``The architectural implications of autonomous driving: Constraints and
  acceleration,'' in \emph{International Conference on Architectural Support
  for Programming Languages and Operating Systems (ASPLOS)}, 2018.

\bibitem{ros-msg}
``Ros messages,'' \url{http://wiki.ros.org/Messages/}, 2020.

\bibitem{rosbot}
``Rosbot 2.0 pro,''
  \url{https://store.husarion.com/collections/dev-kits/products/rosbot-pro/},
  2020.

\bibitem{app-autorace}
``Turtlebot3 autorace,''
  \url{https://emanual.robotis.com/docs/en/platform/turtlebot3/autonomous_driving},
  2020.

\bibitem{turtlebot3}
``Turtlebot3,''
  \url{https://emanual.robotis.com/docs/en/platform/turtlebot3/overview/},
  2020.

\bibitem{gazebo}
``Gazebo 3d robot simulator.'' \url{http://gazebosim.org/}, 2020.

\bibitem{lgsvl}
``Lgsvl simulator.'' \url{https://www.lgsvlsimulator.com/}, 2020.

\bibitem{rviz}
``Rviz 3d visualization tool for ros.''
  \url{https://www.stereolabs.com/docs/ros/rviz/}, 2020.

\bibitem{app-teleop}
``Rosbot teleoperation app,''
  \url{https://husarion.com/tutorials/ros-tutorials/3-simple-kinematics-for-mobile-robot/},
  2020.

\bibitem{app-voice}
``Xiaoqiang voice interaction app,''
  \url{https://community.bwbot.org/topic/492/}, 2020.

\bibitem{app-mapping}
``Rosbot slam app,''
  \url{https://husarion.com/tutorials/ros-tutorials/6-slam-navigation/}, 2020.

\bibitem{app-navi}
``Rosbot navigation app,''
  \url{https://husarion.com/tutorials/ros-tutorials/7-path-planning/}, 2020.

\bibitem{app-explore}
``Rosbot exploration app,''
  \url{https://husarion.com/tutorials/ros-tutorials/8-unknown-environment-exploration/},
  2020.

\bibitem{mobileye}
S.~Shalev-Shwartz, S.~Shammah, and A.~Shashua, ``Reinforcement learning for
  autonomous driving,'' in \emph{NIPS Workshop on Learning, Inference and
  Control of Multi-Agent Systems}, 2016.

\bibitem{udacity}
``An open source self-driving car,'' https://www.udacity.com/self-driving-car/,
  2020.

\bibitem{buf-overflow-1}
B.~B. Madan, M.~Banik, and D.~Bein, ``Securing unmanned autonomous systems from
  cyber threats,'' \emph{Journal of Defense Modeling \& Simulation}, vol.~16,
  no.~2, pp. 119--135, 2016.

\bibitem{buf-overflow-2}
M.~Hooper, Y.~Tian, R.~Zhou, B.~Cao, A.~P. Lauf, L.~Watkins, W.~H. Robinson,
  and W.~Alexis, ``A review on cybersecurity vulnerabilities for unmanned
  aerial vehicles,'' in \emph{Military Communications Conference (MILCOM)},
  2016.

\bibitem{buf-overflow-3}
C.~G.~L. Krishna and R.~R. Murphy, ``A review on cybersecurity vulnerabilities
  for unmanned aerial vehicles,'' in \emph{IEEE International Symposium on
  Safety, Security, and Rescue Robotics (SSRR)}, 2017.

\bibitem{input-bugs}
T.~Kim, C.~H. Kim, J.~Rhee, F.~Fei, Z.~Tu, G.~Walkup, X.~Zhang, X.~Deng, and
  D.~Xu, ``Rvfuzzer: Finding input validation bugs in robotic vehicles through
  control-guided testing,'' in \emph{USENIX Security Symposium (USENIX Security
  19)}, 2019.

\bibitem{ifttt-1}
C.~Nandi and M.~D. Ernst, ``Automatic trigger generation for rule-based smart
  homes,'' in \emph{PLAS@CCS}, 2016.

\bibitem{ifttt-2}
T.~Yu, V.~Sekar, S.~Seshan, Y.~Agarwal, and C.~Xu, ``Handling a trillion
  (unfixable) flaws on a billion devices: Rethinking network security for the
  internet-of-things,'' in \emph{HotNets}, 2015.

\end{thebibliography}
